\documentclass[a4paper,12pt]{article}
\usepackage{cite}
\usepackage{epsfig}
\usepackage{amssymb,amsmath}
\usepackage{graphicx}
\usepackage{float}
\usepackage{comment}
\usepackage{epstopdf}
\usepackage{soul}

\pdfoutput=1

\newlength{\dinwidth}
\newlength{\dinmargin}
\newcommand{\dof}{{ \mathrm{dof}  }}

\newcommand{\bel}[1]{\be\label{#1}}
\newcommand{\e}{\mathrm{e}}
\newcommand{\gsim}{~{}_{\textstyle\sim}^{\textstyle >}~}

\newcommand{\ta}[0]{\tilde \alpha}

\usepackage{color}
\definecolor{DarkGreen}{rgb}{0.3,0.5,0.35}
\definecolor{Violet}{rgb}{0.5,0,1}
\definecolor{Brown}{rgb}{0.5,0.25,0}

\def\be{\begin{equation}}
\def\ee{\end{equation}}
\def\beqn{\begin{eqnarray}}
\def\eeqn{\end{eqnarray}}

\def\ba{\begin{array}{c}}
\def\bat{\begin{array}{cc}}
\def\ea{\end{array}}
\def\bi{\begin{itemize}}
\def\ei{\end{itemize}}

\def\be{\begin{equation}}
\def\ee{\end{equation}}

\usepackage{color}
\begin{document}

\begin{flushright}\vbox{
FTUV/13-1029 \\ IFIC/13-78}
\end{flushright}\vskip 30pt

\begin{center}
{\huge\bf Towards a general analysis of LHC data \\[20pt] within two-Higgs-doublet models}
\bigskip\bigskip\bigskip

{\large Alejandro Celis, Victor Ilisie and Antonio Pich}\\[20pt]
{\large Departament de F\'{\i}sica Te\`orica, IFIC, Universitat de Val\`encia -- CSIC\\ Apt. Correus 22085, E-46071 Val\`encia, Spain}
\end{center}
\bigskip \bigskip\bigskip

\begin{abstract}
\noindent
The data accumulated so far confirm the Higgs-like nature of the new boson discovered at the LHC. The Standard Model Higgs hypothesis is compatible with the collider results and no significant deviations from the Standard Model have been observed neither in the flavour sector nor in electroweak precision observables.
We update the LHC and Tevatron constraints on CP-conserving two-Higgs-doublet models without tree-level flavour-changing neutral currents.   While the relative sign between the top Yukawa and the gauge coupling of the $126$~GeV Higgs is found be the same as in the SM, at $90\%$~CL, there is a sign degeneracy in the determination of its bottom and tau Yukawa couplings.  This results in several disjoint allowed regions in the parameter space.
We show how generic sum rules governing the scalar couplings determine the properties of the additional Higgs bosons in the different allowed regions. The role of electroweak precision observables, low-energy flavour constraints and LHC searches for additional scalars to further restrict the available parameter space is also discussed.
\end{abstract}

\newpage

\section{Introduction}
\label{sec:intro}
Experimental data from the ATLAS \cite{Aad:2012tfa,Aad:2013wqa}, CMS \cite{Chatrchyan:2012ufa,Chatrchyan:2013lba}, D\O~and CDF \cite{Aaltonen:2012qt} collaborations confirm that the new boson discovered at the LHC is related to the mechanism of electroweak symmetry breaking.  The masses of the new boson measured by ATLAS ($125.5 \pm 0.2  ~^{+0.5}_{-0.6} $~GeV) and CMS ($125.7 \pm 0.3 \pm 0.3  $~GeV) are in good agreement, giving the average value $M_h = 125.64\pm 0.35$~GeV, and its spin/parity is compatible with the Standard Model (SM) Higgs boson hypothesis, $J^{P} = 0^{+}$~\cite{Aad:2013xqa,Chatrchyan:2012jja,D0-6387}.  Global analyses of current data find to a good accuracy that the new $h(126)$ boson couples to the vector bosons $(W^{\pm}, Z)$ with the required strength to restore perturbative unitarity in vector boson scattering amplitudes.  The $h(126)$ couplings to fermions of the third generation are also found to be compatible with the SM Higgs scenario~\cite{Cheung:2013kla,LHCP2013}.

A complex scalar field transforming as a doublet under $\mathrm{SU}(2)_{\mathrm{L}}$ seems at present the most elegant and simple explanation for elementary particle masses.  None of the fundamental principles of the SM, however, forbids the possibility that a richer scalar sector is responsible for the electroweak symmetry breaking.  Unlike the addition of new fermion generations or new gauge bosons, an enlarged scalar sector remains in general much more elusive to experimental constraints.  Two-Higgs-doublet models (2HDMs) provide a minimal extension of the SM scalar sector that naturally accommodates the electroweak precision tests, giving rise at the same time to many interesting phenomenological effects~\cite{Gunion:1989we}. The scalar spectrum of a two-Higgs-doublet model consists of three neutral and one charged Higgs bosons.  The direct search for additional scalar states at the LHC or indirectly via precision flavour experiments will therefore continue being an important task in the following years.

Many analyses of LHC and Tevatron data have been performed recently within the framework of CP-conserving 2HDMs with natural flavour conservation (NFC)~\cite{Barroso:2013zxa,Grinstein:2013npa,Eberhardt:2013uba,Chen:2013rba,Craig:2013hca,Coleppa:2013dya,Shu:2013uua,Chiang:2013ixa,Krawczyk:2013jta,Goudelis:2013uca,Arhrib:2013ela,Belanger:2013xza,Enberg:2013ara,WoudaII,Chang:2013ona,Cheung:2013rva
}. These works have focused on different versions of the 2HDM in which a discrete $\mathcal{Z}_2$ symmetry is imposed in the Lagrangian to eliminate tree-level flavour-changing neutral currents (FCNCs).   A more general alternative is to assume the alignment in flavour space of the Yukawa matrices for each type of right-handed fermion~\cite{Pich:2009sp}.   The so-called aligned two-Higgs-doublet model (A2HDM) contains as particular cases the different versions of the 2HDM with NFC, while at the same time introduces new sources of CP violation beyond the CKM phase. First studies of the $h(126)$ boson data within the A2HDM, in the CP-conserving limit, were performed in Refs.~\cite{Cervero:2012cx,Altmannshofer:2012ar,Bai:2012ex
,Celis:2013rcs} and more recently in Refs.~\cite{Barger:2013ofa,Lopez-Val:2013yba,Ilisie:2013cxa}.  The implications of new sources of CP violation within this model for the $h(126)$ phenomenology were also analyzed in Ref.~\cite{Celis:2013rcs}.

In this work we extend the analysis of Ref.~\cite{Celis:2013rcs} and
update the bounds that current LHC and Tevatron data impose on the CP-conserving A2HDM, taking into account the latest results released by the experimental collaborations after the first LHC shutdown.  We also discuss the role of electroweak precision observables and flavour constraints to further restrict the parameter space.
The allowed regions are classified according to the sign of the bottom and tau Yukawa couplings of the $h(126)$ boson, relative to its coupling to vector bosons.   Due to generic sum rules governing the scalar couplings~\cite{Gunion:1990kf,Grzadkowski:1999wj,Ginzburg:2004vp,Celis:2013rcs}, the properties of the additional scalar fields of the model are very different in each of these allowed regions.    We consider also current limits from the search of additional scalars at the LHC and its impact on our knowledge of the $h(126)$ properties. The possibility of a fermiophobic charged Higgs~\cite{Celis:2013rcs} is also analyzed in light of the latest LHC data.   A study of CP-violating effects in the 2HDM along the lines of Ref.~\cite{Celis:2013rcs}  will be deferred to a future work.

This paper is organized as follows.  The present bounds from LHC and Tevatron data are analyzed in section~\ref{sec:CPconserving}, discussing also the role of the loop-induced processes $Z \rightarrow \bar b b$ and $\bar B \rightarrow X_s \gamma$ to further constrain the available parameter space.   In section~\ref{sec:spectra} we consider the search for additional Higgs bosons at the LHC.   The particular case of a fermiophobic charged Higgs is analyzed in section~\ref{sec:fermiophobic}.    A comparison of our findings with those of related works is done in section~\ref{compas} and a summary of our results is finally given in section~\ref{sec:summary}.

\section{A2HDM fit in the CP-conserving limit}
\label{sec:CPconserving}
Let us consider the scalar sector of the CP-conserving 2HDM.  In the so-called Higgs basis where only one of the doublets acquires a vacuum expectation value, the two doublets are parametrized as~\cite{Celis:2013rcs}
\begin{align}  \label{Higgsbasis}
\Phi_1 =\left[ \begin{array}{c}  G^+ \\ \frac{1}{\sqrt{2}}\, (v + S_1 + i G^0) \end{array} \right]  \!\  ,  & &
\Phi_2 = \left[ \begin{array}{c}  H^+ \\ \frac{1}{\sqrt{2}}\, (S_2 + i S_3)   \end{array}\right]  \!\ .
\end{align}
Thus, $\Phi_1$ plays the role of the SM scalar doublet with
$v  = (\sqrt{2}\, G_F)^{-1/2} \simeq 246~\mathrm{GeV}$. The physical scalar spectrum consists of five degrees of freedom: two charged fields $H^\pm(x)$
and three neutral scalars $\varphi_i^0(x)=\{h(x),H(x),A(x)\}$. The later are related with the $S_i$ fields through an orthogonal transformation $\varphi^0_i(x) = \mathcal{R}_{ij} S_j(x)$, which is determined by the scalar potential~\cite{Celis:2013rcs}. In the most general case, the CP-odd component $S_3$ mixes with the CP-even fields
$S_{1,2}$ and the resulting mass eigenstates do not have definite CP quantum numbers.  For a CP-conserving potential this admixture disappears, giving $A(x) = S_3(x)$ and\footnote{In a generic scalar basis $\phi_a(x)$ ($a=1,2$) in which both doublets acquire vacuum expectation values: $\langle 0|\phi_a^T(x)|0\rangle =\frac{1}{\sqrt{2}}\, (0,v_a\, \e^{i\theta_a})$, we have $\tilde \alpha = \alpha - \beta$ in the usually adopted notation.
The angle $\alpha$ determines $h$ and $H$ in terms of the CP-even fields and $\tan \beta = v_2/v_1$ is the ratio of vacuum expectation values.       Given that the choice of basis is arbitrary, the parameters $\alpha$ and $\beta$ are in general unphysical.  These angles are meaningful only in particular models in which a specific basis is singled out (through a symmetry for example)~\cite{Davidson:2005cw}.  }
\bel{eq:mixing}
\left(\ba h \\  H\ea\right)\; = \;
\left[\bat  \cos{\tilde\alpha} &  \sin{\tilde\alpha} \\  -\sin{\tilde\alpha} &  \cos{\tilde\alpha} \ea\right]\;
\left(\ba S_1\\  S_2\ea\right) \, .
\ee
Performing a phase redefinition of the neutral CP-even fields, it is possible to fix the sign of $\sin{\ta}$.  In this work we adopt the conventions\ $M_h \le M_H$\ and\
$ 0 \leq \ta \leq \pi$, so that $\sin{\ta}$ is always positive. To avoid FCNCs, we assume the alignment in flavour space of the Yukawa matrices.   In terms of the fermion mass-eigenstate fields, the Yukawa interactions of the A2HDM read~\cite{Pich:2009sp}
\beqn\label{lagrangian}
 \mathcal L_Y & = &  - \frac{\sqrt{2}}{v}\; H^+ \left\{ \bar{u} \left[ \varsigma_d\, V M_d \mathcal P_R - \varsigma_u\, M_u V \mathcal P_L \right]  d\, + \, \varsigma_l\, \bar{\nu} M_l \mathcal P_R l \right\}
\nonumber \\
& & -\,\frac{1}{v}\; \sum_{\varphi^0_i, f}\, y^{\varphi^0_i}_f\, \varphi^0_i  \; \left[\bar{f}\,  M_f \mathcal P_R  f\right]
\;  + \;\mathrm{h.c.} \, ,
\eeqn
where $\mathcal P_{R,L}\equiv \frac{1\pm \gamma_5}{2}$ are the right-handed and left-handed chirality projectors, $M_f$ the diagonal fermion mass matrices
and $\varsigma_f$ ($f=u,d,l$) the family-universal alignment parameters.
The only source of flavour-changing phenomena is the CKM matrix $V$.
The well-known versions of the 2HDM with NFC are recovered as particular limits of this parametrization, given in Table~\ref{tab:modelsZ2}.

In the present analysis we neglect possible CP-violating effects; {\it i.e.}, we consider a CP-conserving scalar potential and real alignment parameters $\varsigma_f$.
The couplings of the neutral scalar fields are then given, in units of the SM Higgs couplings, by
\begin{align}  \label{equations1}
&& y_{f}^h & = \cos{\tilde\alpha} + \varsigma_f \sin{\tilde \alpha} \!\ , &&& y_{d,l}^A & =  i\,\varsigma_{d,l}  \!\ , &&  \notag \\
& & y_{f}^H & = -\sin{\tilde\alpha} + \varsigma_f \cos{\tilde \alpha} \!\ , &&&
y_{u}^A \; & =\; -i\, \varsigma_u  \!\  \,,
\end{align}
for the fermionic couplings and ($\kappa_V^{\varphi^0_i}\equiv
g_{\varphi^0_iVV}/g_{hVV}^{\mathrm{SM}}$, $V=W,Z$)
\be\label{equations2}
\kappa_V^{h}\;=\; \cos{\tilde \alpha} \, , \qquad\qquad
\kappa_V^{H}\;=\; -\sin{\tilde \alpha} \, , \qquad\qquad
\kappa_V^{A}\;=\; 0  \,,
\ee
for the gauge couplings.
The CP symmetry implies a vanishing gauge coupling of the CP-odd scalar.  In the limit $\tilde\alpha\to 0$, the $h$ couplings are identical to those of the SM Higgs field and the heavy CP-even scalar $H$ decouples from the gauge bosons.\footnote{
The scalar mixing is often parametrized in terms of $\alpha'=\tilde{\alpha}+\frac{\pi}{2}$, so that $\kappa_V^h = \sin{\alpha'}$
and the SM limit corresponds to  $\alpha' = \pi/2$~\cite{Gunion:1989we}. We prefer to describe small deviations from the SM limit with $\tilde \alpha \simeq 0$.}

\begin{table}[t]\begin{center}
\caption{ \it \small CP-conserving 2HDMs based on discrete $\mathcal Z_2$ symmetries. }
\vspace{0.2cm}
\begin{tabular}{|c|c|c|c|}
\hline
Model & $\varsigma_d$ & $\varsigma_u$ & $\varsigma_l$  \\
\hline
Type I  & $\cot{\beta}$ &$\cot{\beta}$ & $\cot{\beta}$ \\
Type II & $-\tan{\beta}$ & $\cot{\beta}$ & $-\tan{\beta}$ \\
Type X (lepton-specific) & $\cot{\beta}$ & $\cot{\beta}$ & $-\tan{\beta}$ \\
Type Y (flipped) & $-\tan{\beta}$ & $\cot{\beta}$ & $\cot{\beta}$ \\
Inert  & 0 & 0 & 0 \\
\hline
\end{tabular}
\label{tab:modelsZ2}
\end{center}\end{table}

\begin{figure}[th!]
\centering
\includegraphics[width=7.3cm,height=6.9cm]{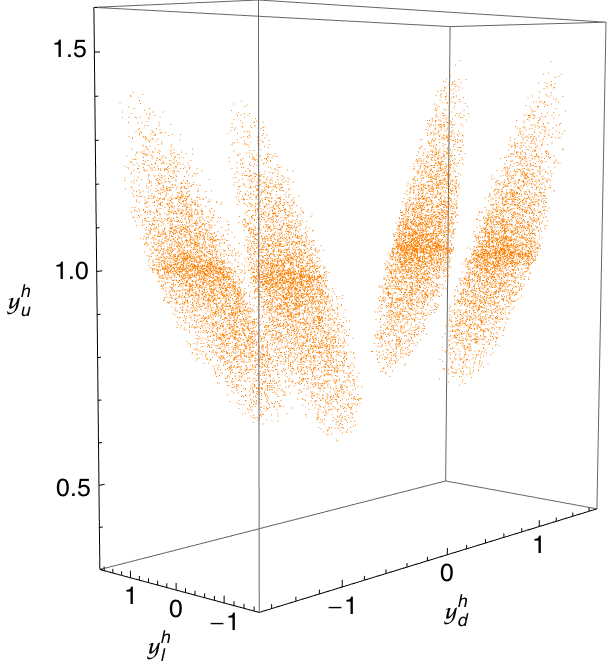}
\hskip 1.2cm
\includegraphics[width=7.3cm,height=7.3cm]{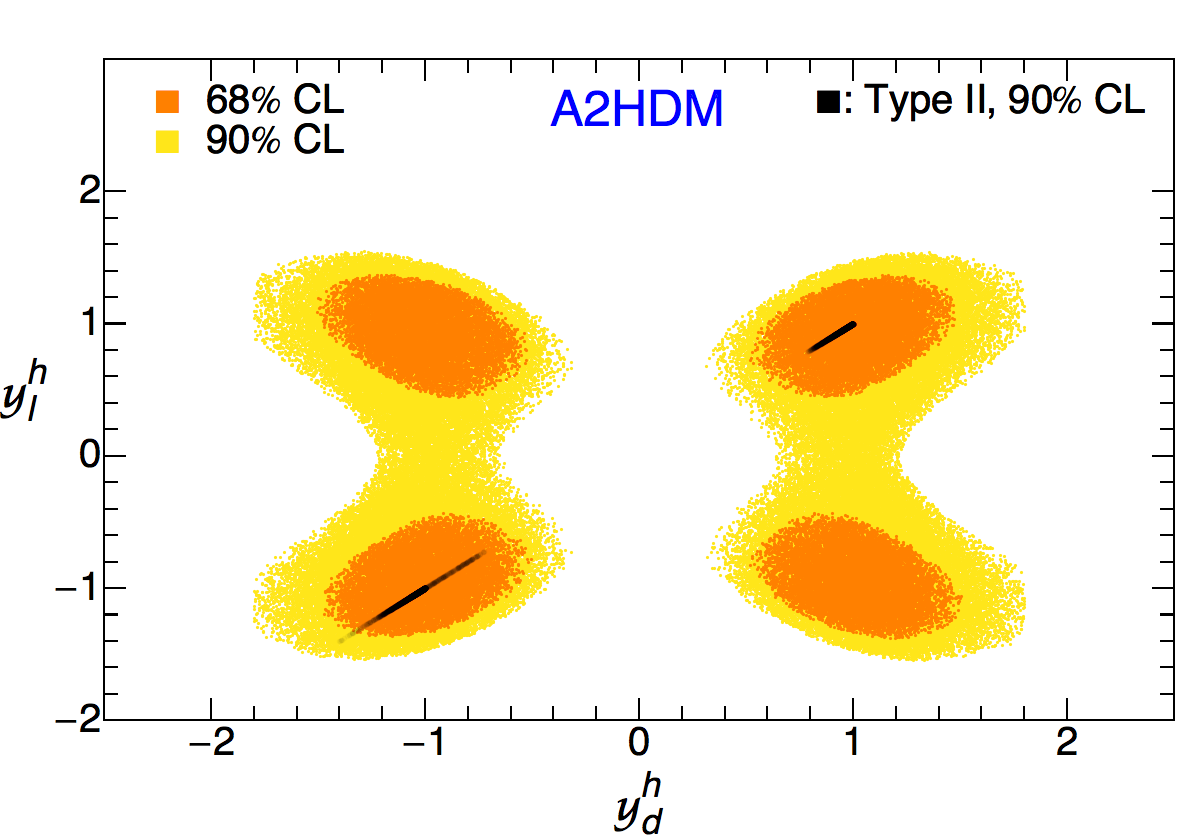}\\
\includegraphics[width=7.3cm,height=7.3cm]{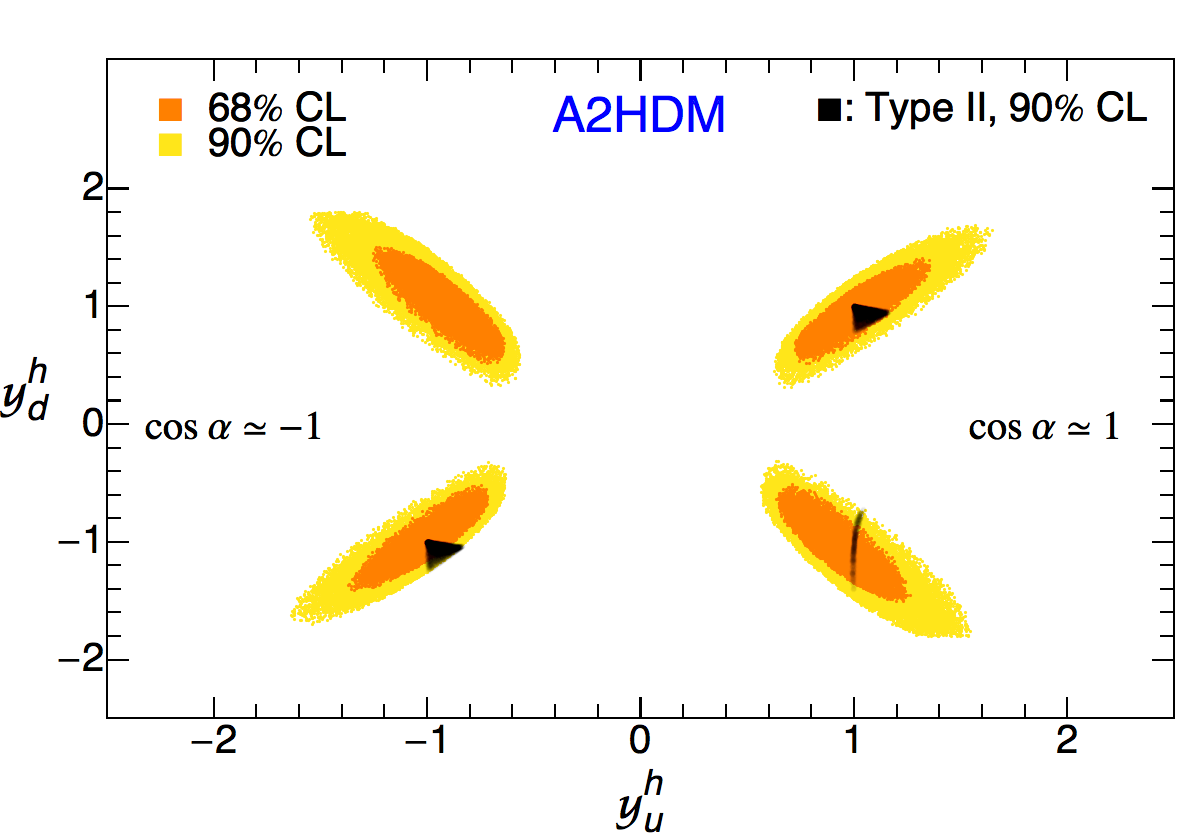}
\hskip 1.2cm
\includegraphics[width=7.3cm,height=7.3cm]{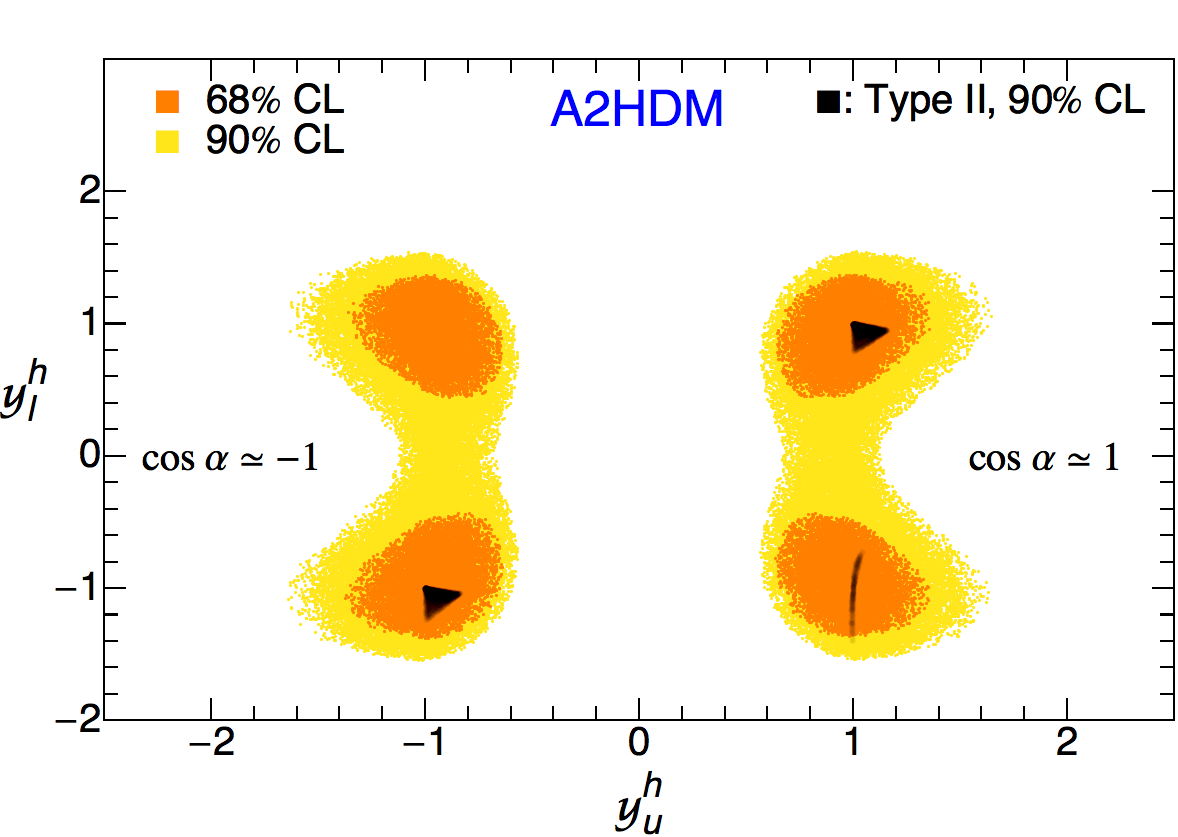}\\
\caption{\label{fig:Reali} \it \small
Allowed regions in the planes $y^h_d - y_l^h$ (top-right),  $y_u^h- y_d^h$ (bottom-left) and $y_u^h- y_l^h$ (bottom-right) at 68\% (orange, dark) and 90\% (yellow, light)~CL from a global fit of LHC and Tevatron data, within the CP-conserving A2HDM.  The particular case of the discrete $\mathcal{Z}_2$ model of type II  is also indicated at 90\% CL (black).   Top-left panel: Allowed region in the space $(y_{u}^{h},y_{d}^{h},y_{l}^{h})$ with $\cos{\tilde \alpha} >0$ at $68\%$~CL (orange).}
\end{figure}

\subsection{Implications of LHC and Tevatron data for the $\mathbf{h(126)}$ boson  \label{LHCandTeV} }
We assume that the $h(126)$ boson corresponds to the lightest CP-even scalar $h$ of the CP-conserving A2HDM.  Current experimental data require its gauge coupling to have a magnitude close to the SM one; {\it i.e.}, $|\cos \tilde \alpha | \sim 1$~\cite{Celis:2013rcs}.   A global fit of the parameters $(\cos \ta, \varsigma_{u}, \varsigma_d, \varsigma_{l})$ to the latest LHC and Tevatron data gives ($\chi_{\mathrm{min}}^2/\dof \simeq 0.73$)
\begin{equation}\label{eq:cos}
|\cos{\tilde \alpha}|\; >\;   0.90  \quad (0.80) \,,
\end{equation}
or equivalently $\sin{\tilde\alpha} <   0.44  \, \,(0.60)$,
at 68\%~CL (90\%~CL). The resulting constraints on the Yukawa couplings of $h$ are shown in Figure~\ref{fig:Reali}. The charged Higgs contribution to the $h \rightarrow \gamma \gamma$ amplitude has been assumed to be negligible in this fit. The global fit determines the relative sign between $y_{u}^{h}$ and $g_{hVV}$ to be the same as in the SM. The flipped sign solution for the top Yukawa coupling, which was preferred before Moriond 2013 due to the observed excess in the $\gamma \gamma$ channel~\cite{Celis:2013rcs}, is ruled out by current data at $90\%$~CL.

The partial decay widths of the Higgs decaying into a pair of fermions are not sensitive to the sign of the Yukawa couplings, $\Gamma(h \rightarrow \bar f f) \propto |y_{f}^{h}|^2$.    The loop-induced processes $h \rightarrow \gamma \gamma$ and $gg \rightarrow h$, on the other hand, are sensitive in principle to the $y_{f=u,d,l}^{h}$ signs.  The decay widths, normalized to the SM prediction, can be written in terms of the modified Higgs couplings as,
\be\label{eq:h2gamma}
 \frac{\Gamma( h \rightarrow \gamma \gamma)}{\Gamma( h \rightarrow \gamma \gamma)^{\mathrm{SM}}} \; \simeq \; \left(   0.28 \, y_{u}^{h}  - 0.004 \, y_{d}^{h} -0.0035 \, y_{l}^{h}  -1.27 \, \kappa_{V}^{h}  \right)^2  + \left(  0.006 \, y_{d}^{h} + 0.003 \, y_{l}^{h} \right)^2   \,,
\ee
where we have neglected a possible charged Higgs contribution to $h \rightarrow 2 \gamma$, and
\be\label{eq:h2gluon}
 \frac{\Gamma( h \rightarrow g g)}{\Gamma( h \rightarrow g g)^{\mathrm{SM}}} \; \simeq \; \left(   1.06 \, y_{u}^{h}  - 0.06 \, y_{d}^{h} \right)^2  + \left(  0.09 \, y_{d}^{h}  \right)^2   \,.
\ee
The last terms in (\ref{eq:h2gamma}) and (\ref{eq:h2gluon}) are the absorptive contributions from $\tau^+\tau^-$ and $b\bar b$ loops.  Neglecting the charged Higgs contribution to $h \rightarrow \gamma \gamma$ is well justified if the charged Higgs is very heavy and/or if the cubic Higgs self-coupling $h H^+H^-$ is very small.
Due to their small masses, the tau and bottom contributions are very suppressed and, therefore, flipping the sign of $y_{d,l}^{h}$ has only a very small effect on the relevant partial widths.

The top-left panel in Figure~\ref{fig:Reali} shows the $68\%$~CL allowed regions in the space $(y_{u}^{h},y_{d}^{h},y_{l}^{h})$ with $\cos{\tilde\alpha} >0$. Four disjoint possibilities can be observed, which can be characterized by the relative signs of $y_{d,l}^{h}$ to that of $\kappa_V^h$;
four additional, equivalent, solutions are found flipping simultaneously the signs of $y_{f}^{h}$ and $\cos{\tilde\alpha}$.  We restrict in the rest of this work to the solutions with $\cos \tilde \alpha > 0$.
The other panels show the projections in the planes $y^h_d - y_l^h$ (top-right),  $y_u^h- y_d^h$ (bottom-left) and $y_u^h- y_l^h$ (bottom-right), at 68\% (orange, dark) and 90\% (yellow, light)~CL. The sign degeneracy in the determination of the bottom and tau Yukawa couplings from current data is clearly observed.  At $90\%$~CL, the leptonic Yukawa coupling $y_{l}^{h}$ is found to be compatible with zero and therefore only two disjoint islands remain ($y_{d}^{h} <0$ and $y_{d}^{h}>0$).

Figure~\ref{fig:Reali} shows also (small black areas, $\chi_{\mathrm{min}}^2/\dof \simeq 0.65$) the constraints in the particular case of the type II model ($\varsigma_{d,l}^{\phantom{.}} = -1/\varsigma_u^{\phantom{.}} = -\tan{\beta}$), usually assumed in the literature and realized in minimal supersymmetric scenarios.
The allowed regions get considerably reduced in this case.
This illustrates that there is a much wider range of open phenomenological possibilities waiting to be explored.
The only allowed regions in the type II model are those with identical
$y_{d}^{h}$ and $y_{l}^{h}$ couplings, making a straight line with slope $+1$ in the $y^h_d - y_l^h$ plane. The $y_{d,l}^{h} < 0$ region with $\cos{\tilde \alpha} > 0$ requires a relatively large value of $\tan \beta$ to flip the sign of $y_{d,l}^{h}$. Similar arguments can be made for the other types of 2HDMs with NFC.
For instance, in the type I model ($\varsigma_{u,d,l}^{\phantom{.}} = \cot{\beta}$) the allowed regions are straight lines with slope $+1$
in the three $y^h_f - y_{f'}^h$ planes.

In the following we will keep the discussion within the more general framework provided by the A2HDM. In case any of the versions of the 2HDM with NFC turns out to be (approximately) realized in Nature, an analysis of experimental data within the A2HDM would reveal it.

\begin{figure}[tb]
\centering
\includegraphics[width=6.3cm,height=6.3cm]{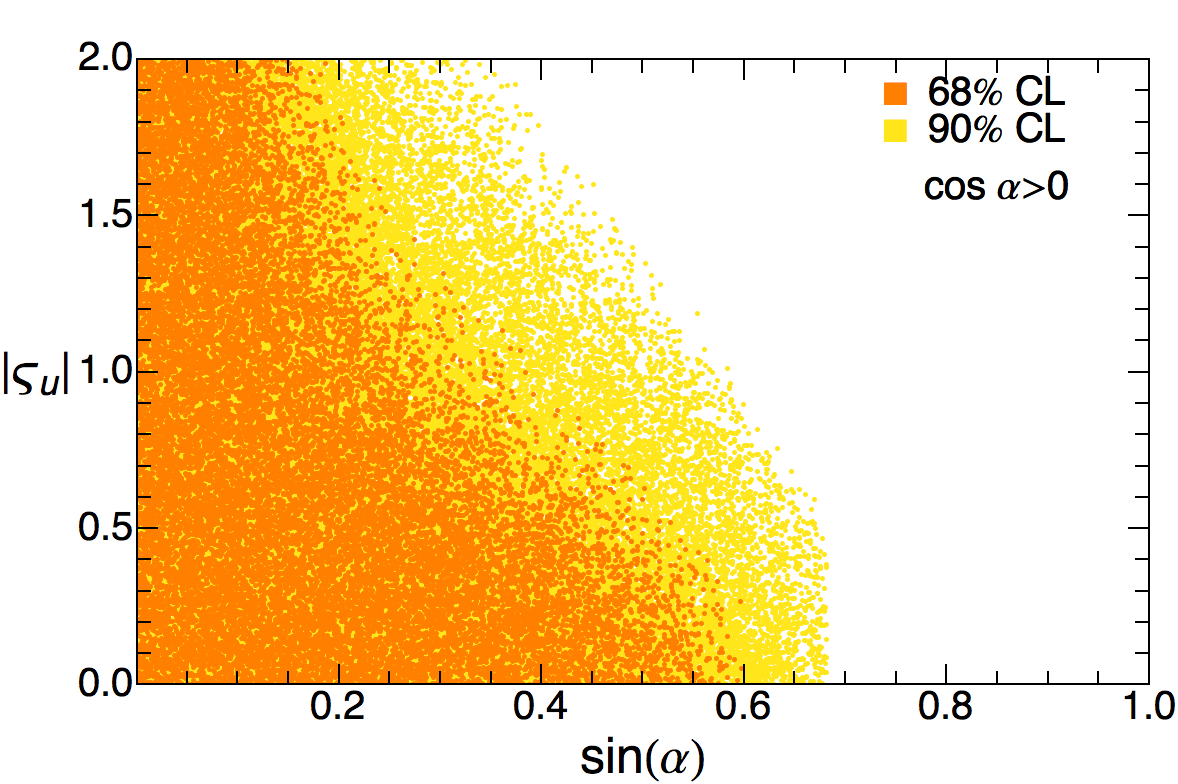}
\caption{\label{fig:varsigma_u} \it \small
Allowed values for $\varsigma_u$, at 68\%~CL (orange) and 90\%~CL (yellow) CL, when $\cos{\tilde \alpha} >0$.}
\end{figure}

\begin{figure}[tb]
\centering
\includegraphics[width=6.3cm,height=6.3cm]{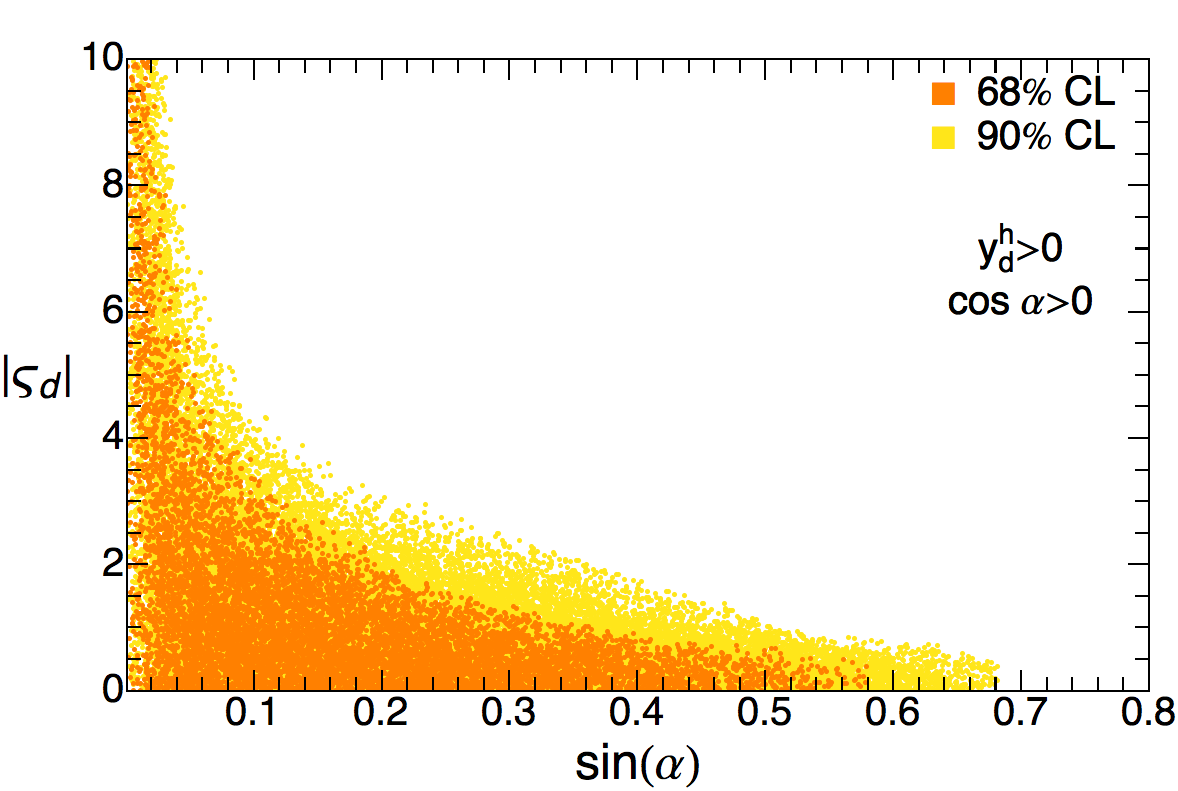}
\hskip 1.5cm
\includegraphics[width=6.3cm,height=6.3cm]{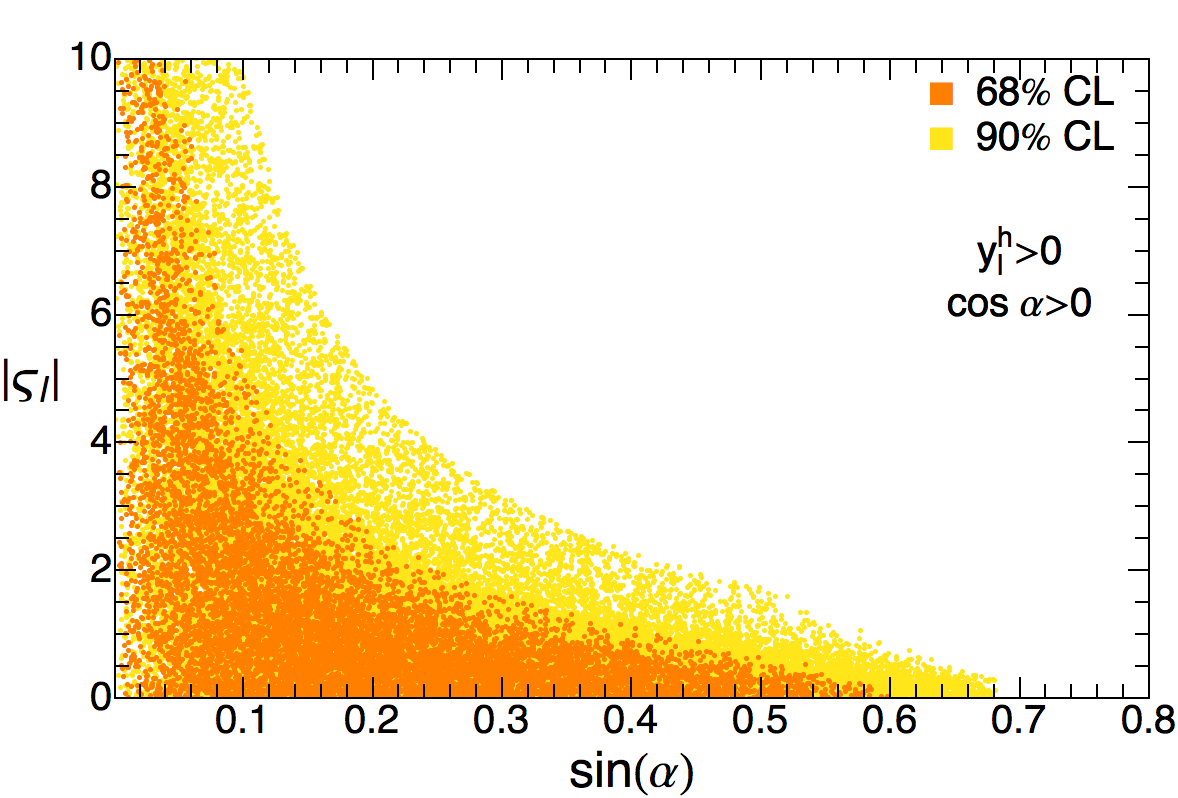}
\caption{\label{fig:RealHdlI} \it \small
Allowed values for $\varsigma_{d,l}$ at 68\%~CL (orange, dark) and 90\%~CL (yellow, light) in the regions where $y_{d}^{h}>0$ (left) or $y_{l}^{h}>0$ (right), keeping only solutions with $\cos{ \tilde \alpha} > 0$. }
\end{figure}

\begin{figure}[tb]
\centering
\includegraphics[width=6.3cm,height=6.3cm]{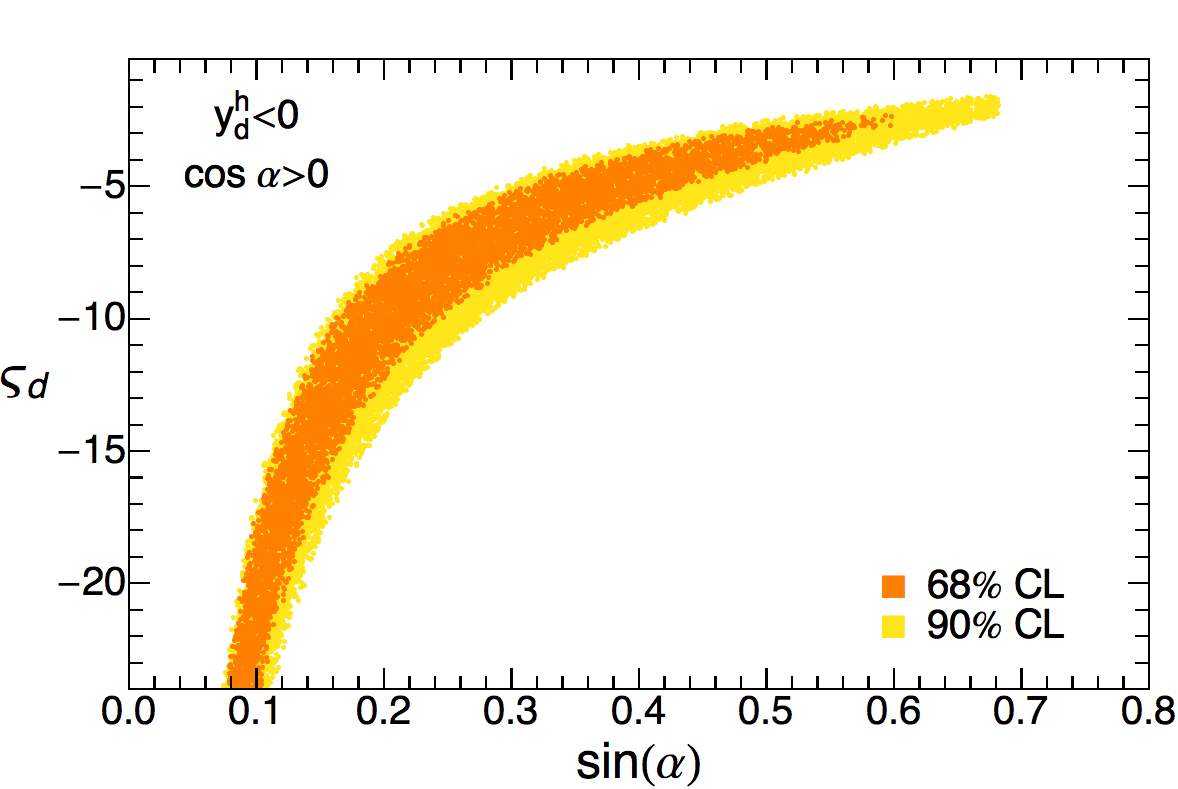}
\hskip 1.5cm
\includegraphics[width=6.3cm,height=6.3cm]{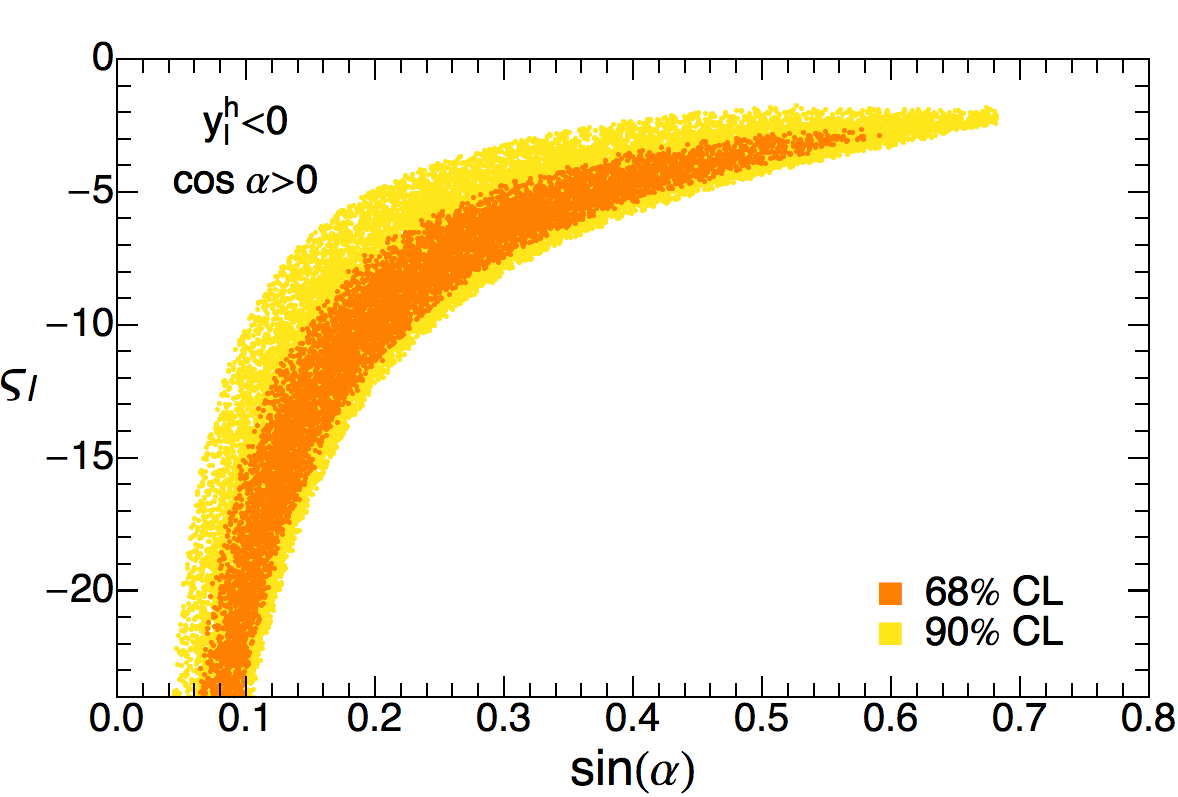}
\caption{\label{fig:RealHdl} \it \small
Allowed values for the alignment parameters $\varsigma_{d,l}$, at 68\%~CL (orange) and 90\%~CL (yellow), in the regions where $y_{d}^{h}<0$ (left) or $y_{l}^{h}<0$ (right), keeping only solutions with $\cos{\tilde \alpha} > 0$. }
\end{figure}

Figures~\ref{fig:varsigma_u}, \ref{fig:RealHdlI} and \ref{fig:RealHdl} show the allowed values for the alignment parameters $\varsigma_{f}$, at 68\% (orange, dark) and $90\%$ (yellow, light) CL, as function of $\sin{\tilde\alpha}$.  Since $y_u^h$ has the same positive sign as $\cos{\tilde\alpha}$ and a similar magnitude, the product $|\varsigma_u| \sin{\tilde\alpha}$ cannot be large. Therefore, $|\varsigma_u|$ gets tightly bounded at large values of $\sin{\tilde\alpha}$ as indicated in Figure \ref{fig:varsigma_u}.   On the other hand, as $\sin{\tilde\alpha}$ approaches zero, all information on $\varsigma_u$ is lost since in this limit the $h$ couplings are SM-like.
The same behaviour is observed in Figure~\ref{fig:RealHdlI}, which shows the allowed values for the alignment parameters  $\varsigma_d$ (left panel) and $\varsigma_l$
(right panel), in the regions with $y_{d}^{h} >0$ or $y_{l}^{h}>0$, respectively.
Important bounds on the magnitudes of $\varsigma_d$ and $\varsigma_l$ are obtained, again, as long as $\sin{ \tilde \alpha} \neq 0$.

A quite different result is obtained in those regions where the Yukawa couplings are negative (again, with $\cos{\tilde\alpha}>0$).
Figure~\ref{fig:RealHdl} shows the allowed values for the alignment parameters $\varsigma_{d, l}$ when $y_{d}^{h}<0$ (left panel) or $y_{l}^{h}<0$ (right panel). A relatively large and negative value for $\varsigma_{d,l}$ is needed to flip the sign in $y_{d,l}^{h}$, given that $\cos{\tilde \alpha} \simeq 1$.
Within the 90\% CL allowed region, $y_{d}^{h} <0$ requires $\varsigma_d \lesssim -2.3$,
while $y_{l}^{h} <0$ implies $ \varsigma_l \lesssim -2.7$.     When $\sin{\tilde \alpha} \lesssim 0.1$, the corresponding values for $|\varsigma_{d,l}|$ become very large: $ \varsigma_{d,l} \lesssim - 24$.

\subsection{SM-like gauge coupling, $\mathbf{\kappa_V^{h} \sim 1}$, without decoupling}
\label{global}
If it is the case that Nature posses an elementary scalar sector composed of two-Higgs doublets, the fact that no large deviations of the $h(126)$ boson properties from the SM have been observed could be pointing towards a decoupling scenario.    In the decoupling limit one of the Higgs doublets can be integrated out, leaving an effective low-energy theory with a SM-like Higgs doublet.  The lightest CP-even Higgs appears with a mass around the electroweak scale and SM-like couplings, while the other scalars are much heavier and degenerate, up to corrections of $\mathcal{O}(v^2)$, $M_{H}^2  \simeq M_A^2 \simeq M_{H^{\pm}}^2 \gg v^2$.   The decoupling limit implies that $|\kappa_{V}^{h}|   \rightarrow 1$, the opposite however is not true. In the limit $|\kappa_{V}^{h}|   \rightarrow 1$, the masses of the additional scalars, $H$, $A$ and $H^{\pm}$, can still be of the order of the electroweak scale~\cite{Gunion:2002zf}.\footnote{  In the Higgs basis~\cite{Celis:2013rcs}, the decoupling limit occurs for $\mu_2 \gg v^2$, where $\mu_2$ is the coefficient of the quadratic $\Phi_2^\dagger\Phi_2$ term in the scalar potential,
while keeping perturbative quartic scalar couplings $ | \lambda_i/4 \pi| \lesssim 1$.   The limit $|\kappa_{V}^{h}|   \rightarrow 1$ without decoupling arises when $\mu_3, \lambda_6 \rightarrow 0$; {\it i.e.}, for vanishing $\Phi_1^\dagger\Phi_2$ and $\Phi_1^\dagger\Phi_1\Phi_1^\dagger\Phi_2$ terms.  For a recent discussion see also Refs.~\cite{Lopez-Val:2013yba,Asner:2013psa,Carena:2013ooa}. }

The decoupling regime is very elusive to experimental tests, leaving a low-energy theory with a light SM-like Higgs, while putting the additional scalars beyond the reach of direct searches at colliders.  Flavour physics constraints are naturally evaded in this case also due to the heaviness of the additional scalars.   Distinguishing signatures of a 2HDM near the decoupling limit would require high-precision measurements of the $h(126)$ boson properties, for example at a future Higgs factory~\cite{Gunion:2002zf}.
In this work, we are interested in the more testable case in which the scalar sector is not in the decoupling regime and all the additional scalars lie around the electroweak scale.    We will assume in particular that the charged Higgs lies in the mass range $M_{H^{\pm}} \in [80, 500]$~GeV.

\begin{figure}[tb]
\centering
\includegraphics[scale=0.24]{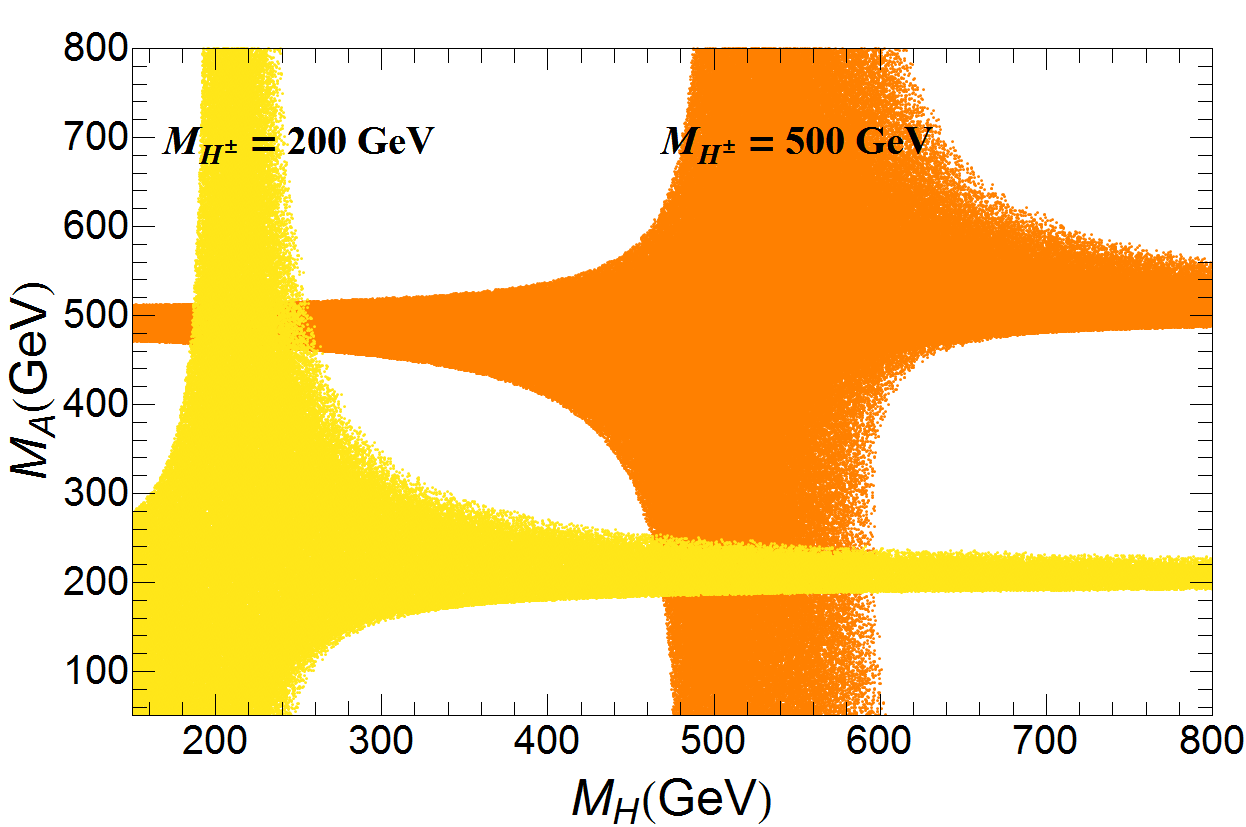}
\caption{\label{fig:oblique} \it \small
Constraints (68\% CL) on the masses of the $H$ and $A$ bosons from the oblique parameters while varying $\cos{\tilde \alpha} \in [0.9,1]$.  The charged Higgs mass is fixed at $M_{H^{\pm}}= 200$~GeV (yellow, light) and $500$~GeV (orange, dark).}
\end{figure}

Deviations from the SM in the gauge-boson self-energies constrain the mass splittings between the additional physical scalars of the 2HDM.
The induced corrections to the oblique parameters have been calculated in Ref.~\cite{He:2001tp} and summarized for the conventions adopted here in Ref.~\cite{Celis:2013rcs}. To satisfy the precision electroweak constraints, the mass differences $| M_{H^{\pm}} - M_H |$ and $| M_{H^{\pm}} - M_A |$ cannot be both large ($\gg v$) at the same time. If there is a light charged Higgs below the TeV scale, an additional neutral boson should be around and vice versa.   Figure~\ref{fig:oblique} shows the $1\sigma$ oblique constraints on the $M_H - M_A$ plane, taking $M_{H^{\pm}}= 200$~GeV (yellow, light) and $500$~GeV (orange, dark), while varying $\cos{\tilde \alpha} \in [0.9,1]$.  The bounds on the mass splittings from the oblique parameters, together with the perturbativity and perturbative unitarity bounds on the quartic-Higgs couplings~\cite{Maalampi:1991fb}, imply that both $H$ and $A$ should have masses below the TeV if $M_{H^{\pm}} < 500$~GeV.  This is the scenario we will be interested in the following, where a rich interplay between precision flavour physics and direct Higgs searches at the LHC can be explored.

\begin{figure}[th!]
\centering
\includegraphics[width=6.cm,height=6.cm]{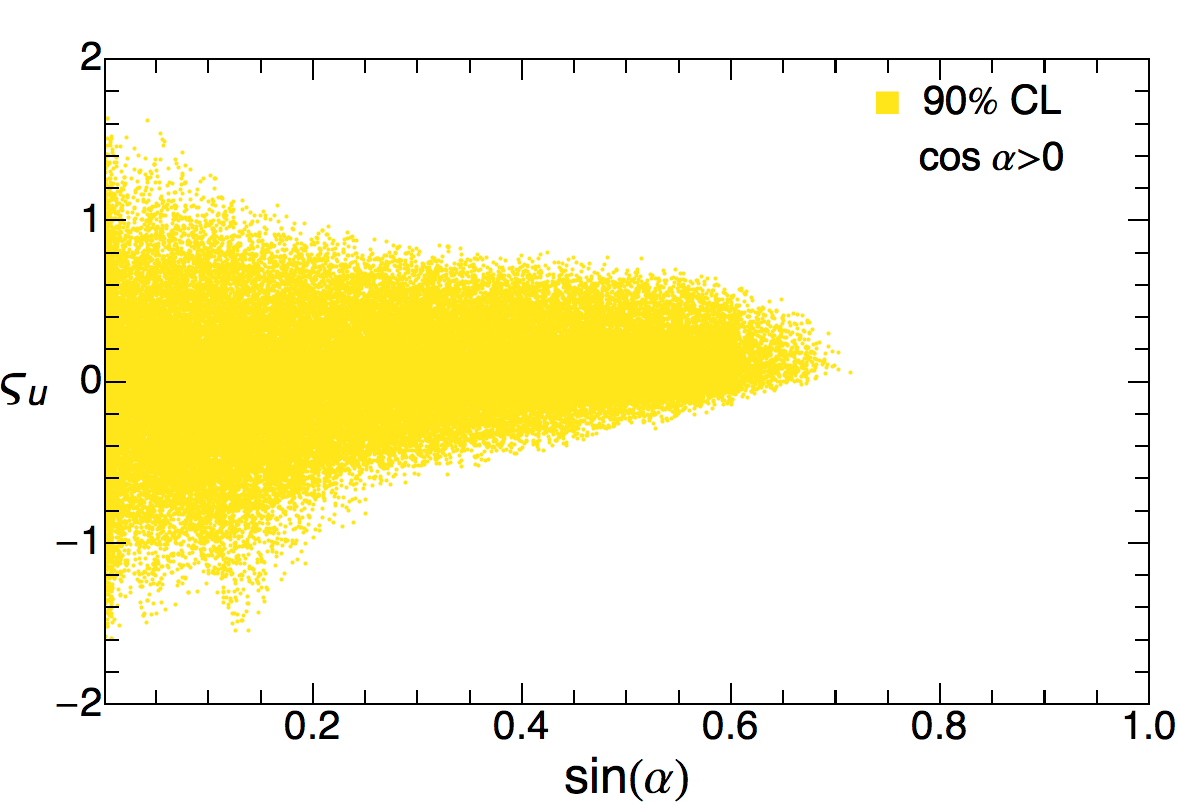}
\hskip 1.2cm
\includegraphics[width=6.cm,height=6.cm]{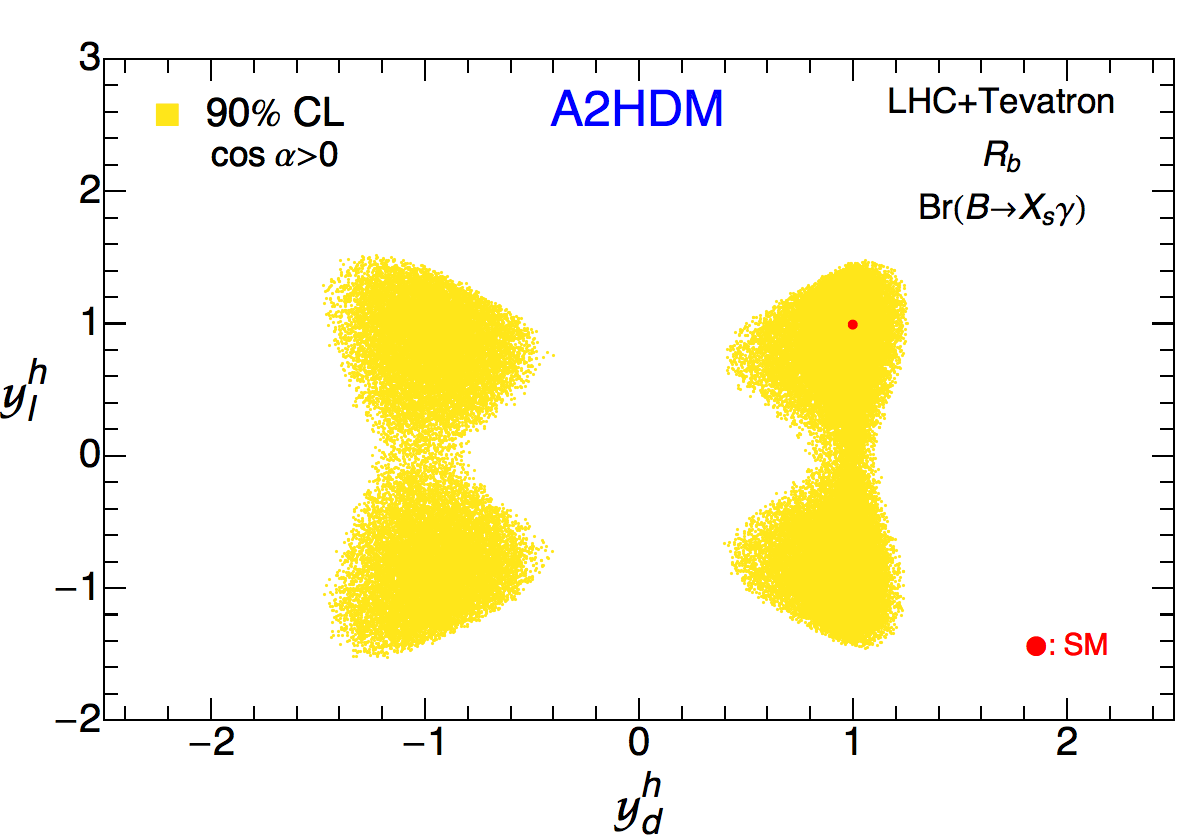}\\
\includegraphics[width=6.cm,height=6.cm]{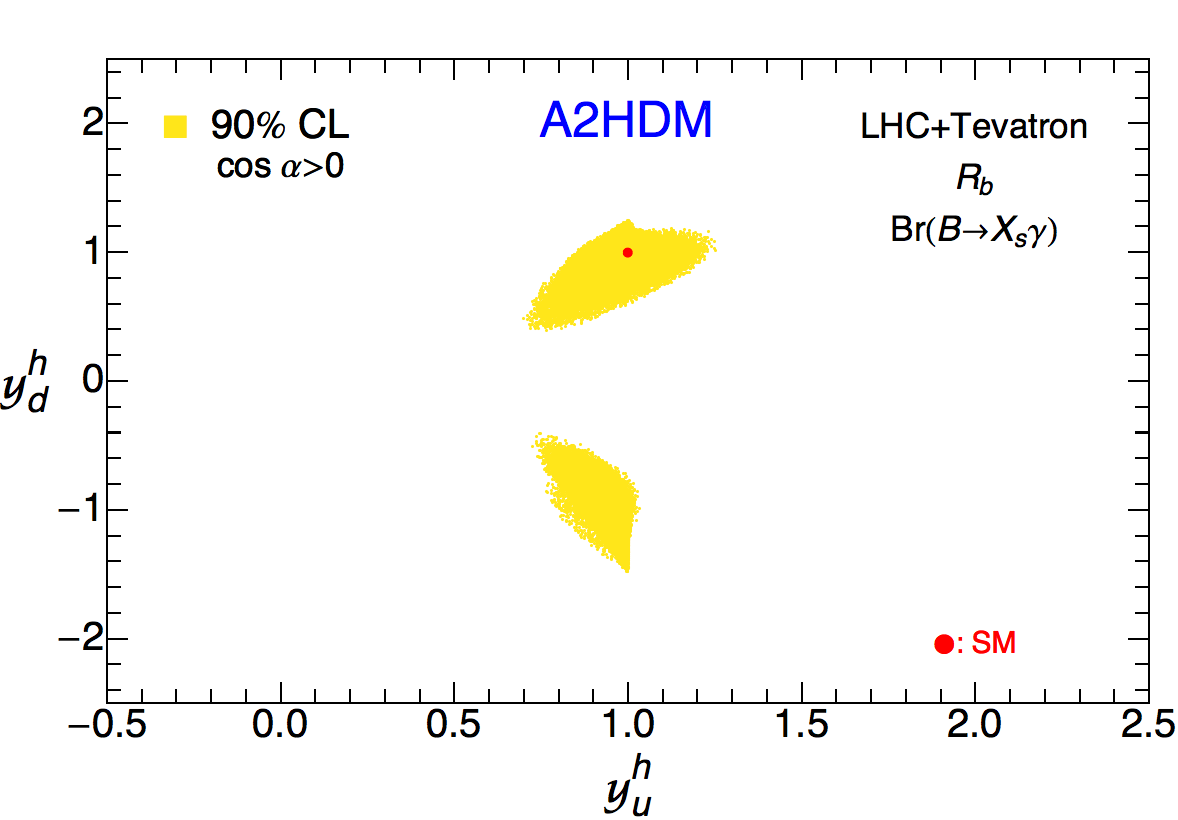}
\hskip 1.2cm
\includegraphics[width=6.cm,height=6.cm]{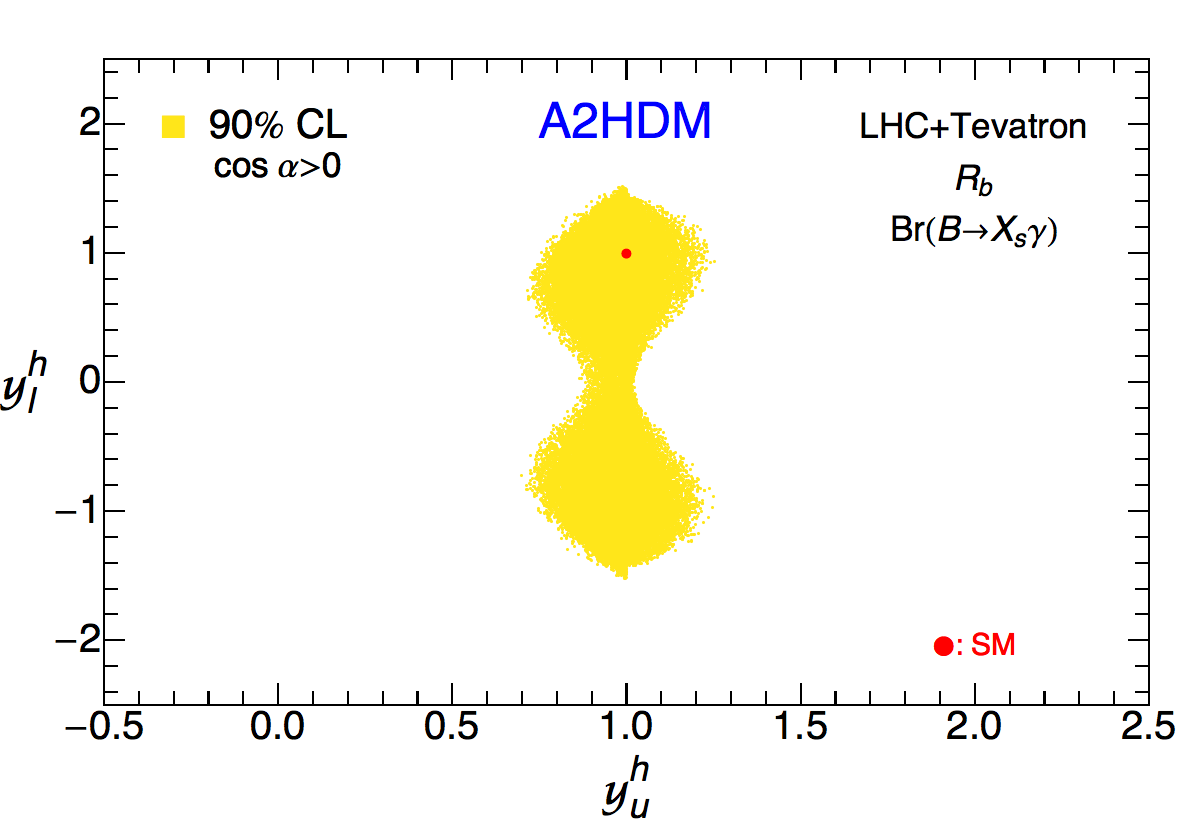}\\
\caption{\label{fig:RealiB} \it \small
Allowed 90\%~CL regions in the planes $\sin \tilde \alpha - \varsigma_u$ (top-left), $y_d^h- y_l^h$ (top-right),  $y_u^h- y_d^h$ (bottom-left), and $y^h_u - y_l^h$ (bottom-right), from a global fit of LHC and Tevatron data together with $R_b$ and $\mathrm{Br}(\bar B \rightarrow X_s \gamma)$, within the CP-conserving A2HDM.  The mass of the charged Higgs is varied within $M_{H^{\pm}} \in [80,500]$~GeV and $\cos{\tilde \alpha} > 0$.   }
\end{figure}

\begin{figure}[th!]
\centering
\includegraphics[width=7.3cm,height=5.3cm]{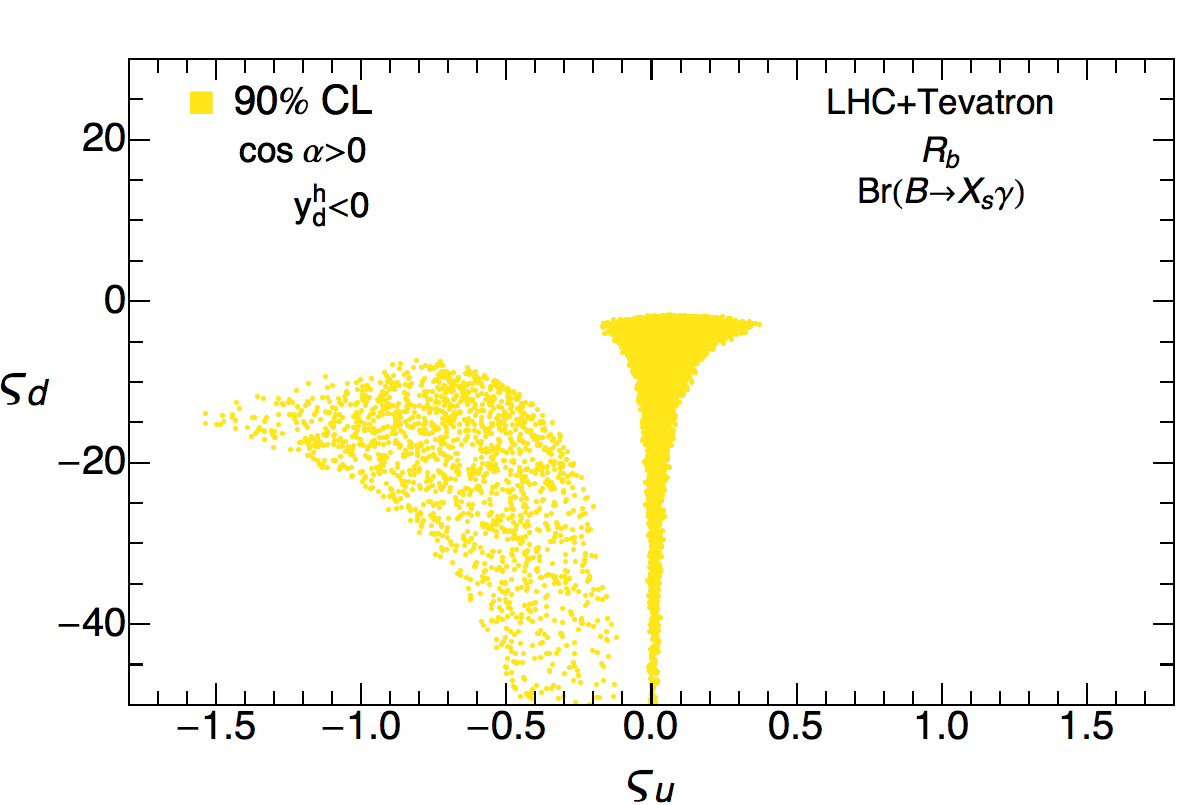}
\hskip 1.2cm
\includegraphics[width=7.3cm,height=5.3cm]{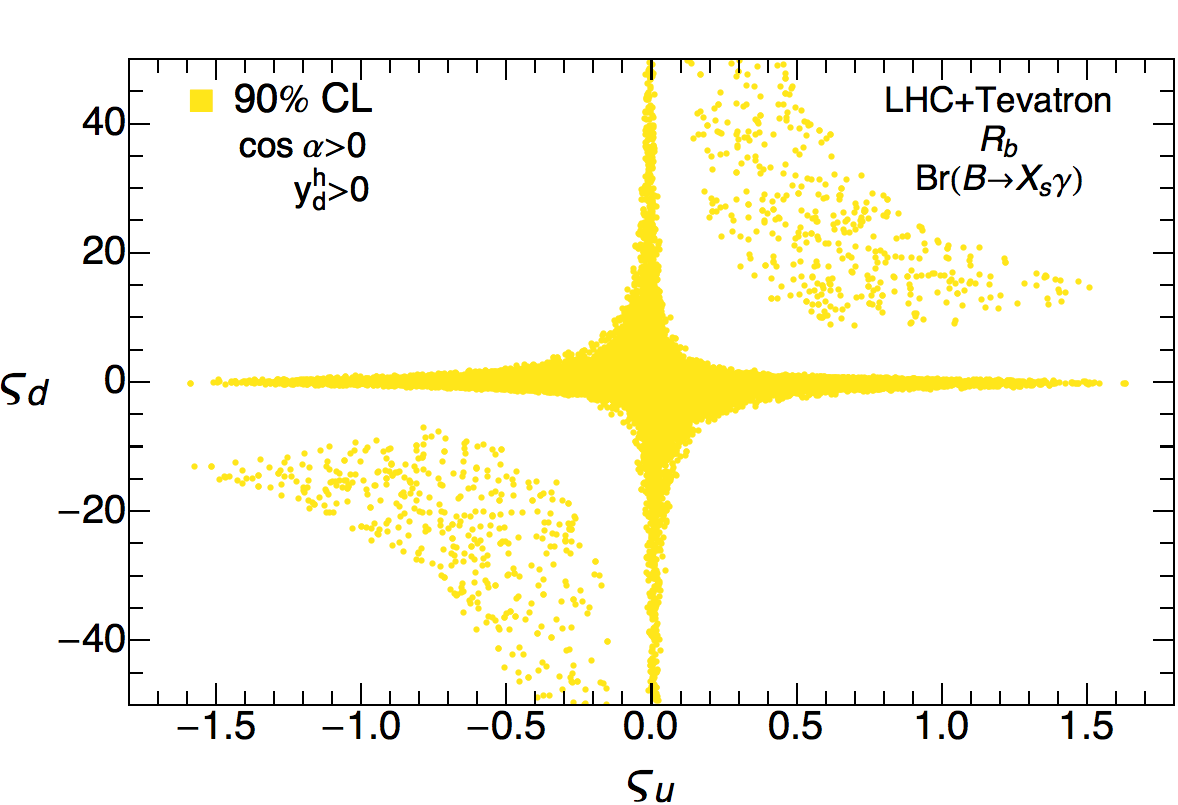}
\caption{\label{fig:fitwflavour} \it \small
Allowed $90\%$~CL region in the plane $\varsigma_u - \varsigma_d$, from LHC and Tevatron data together with $R_b$ and $\mathrm{Br}(\bar B \rightarrow X_s \gamma)$, for $y_{d}^{h}<0$ (left) or $y_{d}^{h}>0$ (right), with $M_{H^{\pm}} \in [80,500]$~GeV and $\cos{\tilde \alpha} > 0$. }
\end{figure}

Interesting constraints are obtained in this case from flavour physics, specially from loop-induced processes with virtual charged Higgs and top quark contributions.
The measured $\bar B^0 - B^0$ mixing and the $Z \rightarrow \bar b b$ decay width require for example that $|\varsigma_u| \lesssim 1.5$, for a charged Higgs below $500$~GeV~\cite{Jung:2010ik}. A more subtle condition can be derived from the radiative decay $\bar{B} \rightarrow X_s \gamma$.  The relevant Wilson coefficients for this process take the form $C_{i}^{\mathrm{eff}} = C_{i, SM} + |\varsigma_u|^2 \, C_{i,uu} - (\varsigma_u^* \varsigma_d) \, C_{i,ud}$, where $C_{i,uu}$ and $C_{i,ud}$ contain the dominant virtual top contributions. Thus, their combined effect can be very different for different values of the ratio $\varsigma_d/\varsigma_u$~\cite{Jung:2010ik,Jung:2010ab,Jung:2012vu}. For real values of the alignment parameters, this provides a very strong bound. For instance, in the type II model, where the two terms interfere constructively, the $\bar{B} \rightarrow X_s \gamma$ rate excludes a charged Higgs mass below
 380~GeV~\cite{Hermann:2012fc}
at $95\%$~CL for any value of $\tan \beta$. In the more general A2HDM framework, a much lighter charged Higgs is still allowed, but in a very restricted region of the parameter space $\varsigma_u-\varsigma_d$~\cite{Jung:2010ik,Jung:2010ab,Jung:2012vu}.

 Semileptonic and leptonic meson decays ($B \rightarrow \tau \nu_{\tau}, D_{(s)} \rightarrow \tau \nu_{\tau} (\mu \nu_{\mu}), B \rightarrow D^{(*)} \tau \nu_{\tau}$), have been analyzed in detail within the A2HDM in Refs.~\cite{Jung:2010ik,Celis:2012dk}.
 These processes put bounds on the combinations $\varsigma_u \varsigma_l/M_{H^{\pm}}^2$ and $\varsigma_d \varsigma_l/M_{H^{\pm}}^2$,
 but the (tree-level) charged Higgs contribution decouples very fast.
 Given that we allow the possibility of a relatively heavy charged Higgs, $M_{H^{\pm}} < 500$~GeV, semileptonic and leptonic decays will not provide complementary information in our analysis.  If one were to focus the discussion to a very light charged Higgs boson, these processes would certainly need to be taken into account.\footnote{The current excess observed by the BaBar collaboration in exclusive $b \rightarrow c \tau \nu$ transitions can only be accommodated within the framework of 2HDMs if one allows for a departure of the Yukawa alignment hypothesis~\cite{Lees:2012xj,Celis:2012dk}.   More theoretical studies on the relevant hadronic matrix elements as well as an update of these modes from the Belle collaboration using the full dataset, are needed to further asses the significance of this excess.}

In Figure~\ref{fig:RealiB} we show the effect of including $\bar{B} \rightarrow X_s \gamma$ and $R_b = \Gamma( Z \rightarrow \bar b b)/\Gamma(Z \rightarrow \text{hadrons})$ in the fit of ($\cos \ta, \varsigma_u, \varsigma_d, \varsigma_l$) while varying $M_{H^{\pm}} \in [80,500]$~GeV and, as usual, keeping only solutions with $\cos \tilde \alpha >0$.   The down-quark and leptonic alignment parameters are varied within $|\varsigma_{d,l} | \leq 50$ to maintain perturbative scalar interactions for bottom quarks and tau leptons.   The charged Higgs contribution to the $2 \gamma$ channel is also neglected in this fit; therefore, $M_{H^{\pm}}$
only enters in the fit through the flavour observables considered.   Strictly, the analysis is then only valid in those regions of the parameter space in which the charged Higgs is reasonably heavy and/or the cubic Higgs self-coupling $h H^+H^-$ is very small.  The results, however, would not change significantly if the $H^\pm$ contribution to $h \rightarrow 2 \gamma$ were included in the fit, since it would be compatible with zero, see section \ref{sec:fermiophobic}.  In the $y_u^{h} - y_{d}^{h}$ plane, it can be observed that a significant part of the previously allowed region is excluded by flavour observables when compared to Figure~\ref{fig:Reali}.   This is due to the effect of $\mathrm{Br}(\bar B \rightarrow X_s \gamma)$ which induces severe constraints in the plane $\varsigma_u-\varsigma_d$, as shown in Figure~\ref{fig:fitwflavour}.

For the case $y_{d}^{h}>0$, collider data do not put any bound on $\varsigma_{u, d}$ in the limit $\sin \tilde \alpha \rightarrow 0$; the only constraint that appears in Figure~\ref{fig:fitwflavour} (right-panel) is therefore coming from $Z \rightarrow \bar b b $ and $\bar{B} \rightarrow X_s \gamma$.  For $y_{d}^{h}<0$, LHC and Tevatron data determine that $\varsigma_d \lesssim - 2$ in order to flip the Yukawa sign, thus excluding a large region that would otherwise be allowed by flavour observables alone.     Compared with Figure~\ref{fig:varsigma_u}, the value of $|\varsigma_u|$ is slightly more constrained by $R_b$;  when $M_{H^{\pm}} < 500$~GeV, one finds $|\varsigma_u| \lesssim 1.5$ for $\sin \tilde \alpha \simeq 0$ while a stronger limit is obtained for larger values of $\sin \tilde \alpha$ due to LHC and Tevatron data.     The corresponding allowed regions shown in Figures~\ref{fig:RealHdlI} and \ref{fig:RealHdl} remain almost identical after adding the flavour observables and, therefore, are not shown here.

\section{Searches for additional Higgs bosons}
\label{sec:spectra}
The search for additional Higgs bosons is one of the most important tasks for the next LHC run.   The current information on the $h(126)$ properties puts relevant constraints on the couplings of the other scalars. In particular, Eqs.~(\ref{equations1}) and (\ref{equations2}) imply the sum rules
\beqn\label{eq:SumRules1}
\left|\kappa_V^H\right|^2 &=& 1 -\left|\kappa_V^h\right|^2\, ,
\\ \label{eq:SumRules2}
\left|y_f^H\right|^2 - \left|y_f^A\right|^2 &=& 1 - \left|y_f^h\right|^2\, ,
\\ \label{eq:SumRules3}
\kappa_V^H \, y_f^H  &=&1 -  \kappa_V^h \,  y_f^h \, .
\eeqn
The first one is just the trivial trigonometric relation between $\sin{\tilde\alpha}$ and $\cos{\tilde\alpha}$, which implies that the gauge coupling $g_{HVV}$ goes to zero when $g_{hVV}$ approaches the SM value.
The lower bound on $|\cos{\tilde\alpha}|$ in Eq.~(\ref{eq:cos}) gives a direct limit on the coupling of the heavy CP-even scalar $H$ to two gauge bosons, with important implications for searches in the $H\to VV$ channels.
The relation (\ref{eq:SumRules2}) constrains the difference of the magnitudes of the $H$ and $A$ Yukawa couplings. When the mixing angle $\tilde\alpha$ becomes zero, $y_f^h=1$ and $\left|y_f^H\right| = \left|y_f^A\right| = \varsigma_f$.  Relation (\ref{eq:SumRules3}) shows that whenever $h$ has a flipped sign Yukawa $(\kappa_V^h \sim 1, y_f^{h} \sim -1)$, the corresponding Yukawa coupling of $H$ must be very large $y_f^H \kappa_V^{H}  \sim 2$.  This sum rule plays a crucial rule in the restoration of perturbative unitarity in $W_L^+ W_L^- \rightarrow f \bar f$ scattering and is behind the particular shape of the allowed regions in Figure \ref{fig:RealHdl}.   The allowed values for $\kappa_V^{h}$ and $y_{f}^{h}$, obtained in section~\ref{global} from $h(126)$ collider data and flavour constraints, imply, due to the sum rules, the following $90\%$~CL bounds:
\beqn\label{eq:SumRulesP}
&&|y_{u}^{H}|^2 -|y_{u}^{A}|^2\in[-0.6,0.5] \,, \qquad\qquad \kappa_{V}^{H} \, y_{u}^{H}\in[-0.17,0.5] \,, \nonumber\\
&&|y_{d}^{H}|^2 -|y_{d}^{A}|^2\in[-1.2,0.9] \,, \qquad\qquad \kappa_{V}^{H} \, y_{d}^{H}\in[-0.3,0.7] \cup [1.3,2.5]\,,  \nonumber \\
&&|y_{l}^{H}|^2 -|y_{l}^{A}|^2\in[-1.3,1.0]  \,, \qquad\qquad  \kappa_{V}^{H} \, y_{l}^{H}\in[-0.5,2.5] \,.\label{Eqfs}
\eeqn

A generic $h(126)$ boson with modified couplings to fermions and gauge bosons would violate perturbative unitarity at high energies, in certain physical processes.
Partial-wave unitarity bounds would be violated for example in  $W_L^+ W_L^- \rightarrow f \bar f$ inelastic scattering at a scale $\sqrt{s} \simeq \Lambda \;=\;  16 \pi v^2/(m_f \, |  1 -  y_{f}^{h} \,  \kappa_{V}^{h}|)$~\cite{Farina:2012xp}. For flipped-sign Yukawa couplings, $\kappa_{V}^{h} \simeq 1$ and $y_{f}^{h} \simeq -1$, we obtain an approximate value of $\Lambda \sim 9$~TeV for the top quark, while $\Lambda \sim 400$~TeV is obtained for the bottom quark and tau lepton due to the fact that they have smaller masses.      A modified $hVV$ coupling would also lead to a violation of perturbative unitarity in $W_L^- W_L^+ \rightarrow W_L^- W_L^+$ elastic scattering; for $\kappa_{V}^{h} = 0.89$ (0.95)  this occurs at a scale $\sqrt{s} = 2.7$ (3.8)~TeV respectively~\cite{Cheung:2008zh}.
The scalar couplings in the 2HDM satisfy generic sum rules which ensure that perturbative unitarity is restored, provided the additional scalar states are light enough.  In the processes considered previously, $W_L^+ W_L^- \rightarrow f \bar f$  and $W_L^- W_L^+ \rightarrow W_L^- W_L^+$, the heavier CP-even Higgs enters with the required couplings to cancel the bad high-energy behavior of the amplitudes.    It must be noted that a given physical state needed to restore perturbative unitarity can appear well below the scale at which the partial-wave unitarity bounds are violated.  This is well known in the SM where the Higgs mass is only weekly bounded by perturbative unitarity: $M_h \lesssim 1$~TeV~\cite{Lee:1977eg}.

The possibility of flipped-sign bottom and/or tau Yukawa couplings has important implications for the properties of the additional Higgs bosons but only subtle effects in the $h(126)$ phenomenology.  Relatively large values for the alignment parameters $\varsigma_{d,l}$ are needed to flip the sign of $y_{d,l}^{h}$ given that $|\kappa_{V}^{h}| \simeq 1$, implying that the additional Higgs bosons $H^{\pm}, H$ and $A$ should posses very large couplings to bottom and/or tau leptons.

The couplings of the missing Higgs bosons $H^\pm$, $H$ and $A$, and therefore their phenomenology, are very different in each of the allowed regions shown in Figure~\ref{fig:Reali}.  It thus seems appropriate to discuss the search strategy for additional scalar states and the experimental constraints in each allowed island separately.
An obvious question to address is how future Higgs searches at the LHC, combined with low-energy precision experiments at the intensity frontier, can be used to
exclude some of the allowed islands and/or determine the right solution chosen by Nature.

The SM-like region with $y_f^h>0$ ($f=u,d,l$) includes the trivial solution $\varsigma_f = 0$. Moreover, the Yukawa couplings $y_{f}^{H}$ are also compatible with zero. Therefore, one has to face the possibility of a SM-like scalar $h$ plus a fermiophobic scalar doublet including the $H$, $A$ and $H^\pm$ fields.
This is a very difficult experimental scenario where the missing scalars decouple from the fermionic sector and also the coupling $g_{HVV}=0$. In this case, the production of the additional scalars can occur for example through the $ZHA$, $ZH^\pm H^\mp$, $W^\pm H^\mp H$ and $W^\pm H^\mp A$ couplings or through the scalar potential.   In the limit $\sin \tilde \alpha = 0$, the $h(126)$ data does not provide any constraints on the alignment parameters $\varsigma_f$ (see Figures~\ref{fig:varsigma_u} and \ref{fig:RealHdlI}). This opens a more interesting possibility with $\left|y_f^H\right| = \left|y_f^A\right| = \varsigma_f$; the $H$ and $A$ bosons could then be produced through the gluon-fusion mechanism or in associated production with a heavy-quark pair. Moreover, since $\varsigma_d$ and $\varsigma_l$ are only weekly constrained by flavour observables, the couplings to bottom quarks and tau leptons could be very sizeable, generating interesting phenomenological signals.   For a very large $|\varsigma_d|$ for example, b-quark associated Higgs production $b \bar b \rightarrow \Phi$ or $g b \rightarrow \Phi b$ can become the dominant production mechanism of the heavy scalars $H$ and $A$ at the LHC.  Similarly, charged Higgs production in association with top and bottom quarks, $gg \rightarrow t \bar b H^-$ or $q \bar q \rightarrow t \bar b H^-$, can be considerably enhanced in this case.  If on the other hand $|\varsigma_l|$ is very large, heavy neutral scalars would probably decay dominantly into leptons, opening the interesting possibility of discovery in the very clean $\Phi \rightarrow \mu^+ \mu^-$ channel.   The charged Higgs also, would be expected to decay dominantly into a $\tau \nu_{\tau}$ pair in this case. 

The situation is rather different in the other three regions with flipped-sign Yukawas: (a) $y_{d}^{h} <0$ and $y_{l}^{h} >0$,  (b) $y_{d}^{h} >0$ and $y_{l}^{h} <0$, and (c) $y_{d,l}^{h} <0$. As shown in Figure~\ref{fig:RealHdl}, the alignment parameters are tightly constrained in these regions and the missing Higgs bosons could have a relatively large coupling to the bottom and/or tau fermions.  In all four allowed regions the alignment parameter $\varsigma_u$ is compatible with zero, therefore there exists the possibility that all production mechanisms of the remaining scalars involving the coupling with top-quarks could be greatly suppressed.

\subsection{Charged Higgs searches} \label{chs}
There are already important exclusion limits coming from charged Higgs searches at colliders, but most of them depend on the assumed Yukawa structure or some hypothesis about the scalar spectrum.
In some cases, however, it is possible to set more general limits.  For instance, a very light charged Higgs would modify the $Z$ boson decay width if the channel $Z \rightarrow H^+ H^-$ is open.  Since the coupling $Z H^+ H^-$ is completely fixed by the gauge symmetry and does not depend on any free parameter of the model, the constraint  $\Gamma_Z^{\mathrm{non\mbox{-}SM}} < 2.9$~MeV (95\% CL) on non-SM decays of the Z boson implies $M_{H^\pm} \gtrsim 39.6$ GeV (95\% CL)~\cite{Abbiendi:2013hk}.
A much stronger lower bound on the $H^\pm$ mass, $M_{H^\pm} \gtrsim 80$ GeV (95\% CL)~\cite{Abbiendi:2013hk}, was set at LEP, assuming that the charged Higgs only decays into $\tau \nu$ or $c s$ final states.    A softer limit would be obtained on the other hand if the $H^+\to W^+ A$ decay is kinematically allowed.  Assuming that $M_A>12$~GeV and a type-I Yukawa structure, the limit $M_{H^\pm} \gtrsim 72.5$ GeV  was obtained in $H^+\to W^+ A\to W^+ b\bar b$ searches~\cite{Abbiendi:2013hk}.

In this section, we consider the LHC searches for a light charged Higgs produced via $t \rightarrow H^+ b$, in the decay channels  $H^+ \rightarrow \tau^+ \nu_{\tau}$~\cite{Aad:2012tj,Chatrchyan:2012vca} and $H^+ \rightarrow c \bar s$~\cite{Aad:2013hla},
which
are kinematically limited to $M_{H^{\pm}} < m_t - m_b$.  We focus on the constraints that can be extracted on the A2HDM from direct charged Higgs searches and flavour observables; the only parameters entering in this analysis are therefore $(M_{H^{\pm}}, \varsigma_u, \varsigma_d, \varsigma_l)$.   A full scan of the A2HDM parameter space, taking into account electroweak precision data, perturbativity and perturbative unitarity bounds, would give as a result that the neutral scalars $H$ and $A$ cannot be arbitrarily heavy and strong correlations in the $M_H - M_A$ plane will appear as those shown in Figure~\ref{fig:oblique}. We refer the reader to appendix~\ref{formulae} for relevant formulae used here.   To a good approximation, the branching ratio for $t \rightarrow H^+ b$ is given by
\be  \label{eq:Charged_decay1}
\mathrm{Br}( t \rightarrow H^+ b   ) \; \simeq\; \dfrac{   \Gamma( t \rightarrow H^+ b  )  }{ \Gamma(  t \rightarrow W^+ b )  +  \Gamma( t \rightarrow H^+ b) } \,,
\ee
where we have neglected CKM-suppressed channels in the total top width.   We do not consider the possibility of a very light CP-odd Higgs boson which could open decay channels like $H^+ \rightarrow W^+ A$; therefore, the charged Higgs decays only into fermions.
Searches into the final state $\tau^+ \nu_{\tau}$ put bounds on the combination
$\mathrm{Br}( t \rightarrow H^+ b  ) \times  \mathrm{Br}(  H^+ \rightarrow \tau^+ \nu  )$,
while current searches for quark decay modes are usually interpreted as limits on $\mathrm{Br}( t \rightarrow H^+ b  )  \times \mathrm{Br}(H^+ \rightarrow c \bar s)$. This is due to the expected dominant decay modes of the charged Higgs in the MSSM scenario or in the type-II 2HDM.  In general, these searches really put bounds on
$\mathrm{Br}( t \rightarrow H^+ b  ) \times  \left[ \mathrm{Br}(  H^+ \rightarrow c \bar s  )  + \mathrm{Br}(  H^+ \rightarrow c \bar b  )    \right]$.
Other final states involving light quarks are neglected as they bring much smaller contributions.

For the next discussion it is useful to write down the following approximate formulae
\beqn  \label{relation1}
\dfrac{  \Gamma(   H^+ \rightarrow c \bar b   ) }{   \Gamma(  H^+ \rightarrow c \bar s  )}  &\simeq &   \dfrac{ |V_{cb} |^2 }{  |V_{cs}|^2   }   \dfrac{  \left(      | \varsigma_d |^2  m_b^2 + |\varsigma_u|^2 m_c^2   \right)      }{ \left(   | \varsigma_d |^2  m_s^2 + |\varsigma_u|^2 m_c^2 \right)   }    \,,  \nonumber \\[5pt]
 \dfrac{ \Gamma(H^+ \rightarrow c \bar b)}{  \Gamma(H^+ \rightarrow \tau^+ \nu_{\tau} ) }  &\simeq &   \dfrac{  N_C | V_{cb}|^2   \left(    | \varsigma_d  |^2  m_b^2 + | \varsigma_u |^2 m_c^2  \right)   }{m_{\tau}^2 | \varsigma_l |^2 } \,.
\eeqn
We can observe that the decay channel $H^+ \rightarrow c \bar b$ can be important, compared with $H^+ \rightarrow c \bar s$, in certain regions of the A2HDM parameter space in which the strong CKM suppression ($ |V_{cb}| \ll |V_{cs}|$) is compensated by a hierarchy of the alignment parameters~\cite{Akeroyd:2012yg}.  Indeed, for $|\varsigma_d | \gg | \varsigma_{u}|, |\varsigma_l|$ the decay channel $H^+ \rightarrow c \bar b$ becomes significant compared with $H^+ \rightarrow c \bar s, \tau^+ \nu_{\tau}$, as shown in Eq.~\eqref{relation1}.  This does not occur in the 2HDMs of types I, II and X, due to correlations between the parameters $\varsigma_{f=u,d,l}$, see Table~\ref{tab:modelsZ2}.    In the type-Y 2HDM, on the other hand, the limit $|\varsigma_d | \gg | \varsigma_{u}|, |\varsigma_l|$ is achieved for large $\tan \beta$; in this case, however, the $\mathrm{Br}(\bar B \rightarrow X_s \gamma)$ constraints forbid a light charged Higgs because $\varsigma_u = - 1/\varsigma_d$~\cite{Akeroyd:2012yg}.
It has been shown in Ref.~\cite{Akeroyd:2012yg} that a dedicated search for $H^+ \rightarrow c \bar b$ decays, implementing a $b$ tag on one of the jets coming from $H^{\pm}$, could provide important constraints on the parameter space region with $|\varsigma_d | \gg | \varsigma_{u}|, |\varsigma_l|$ where this channel becomes important.

\begin{figure}[tb]
\centering
\includegraphics[width=7.3cm,height=6.3cm]{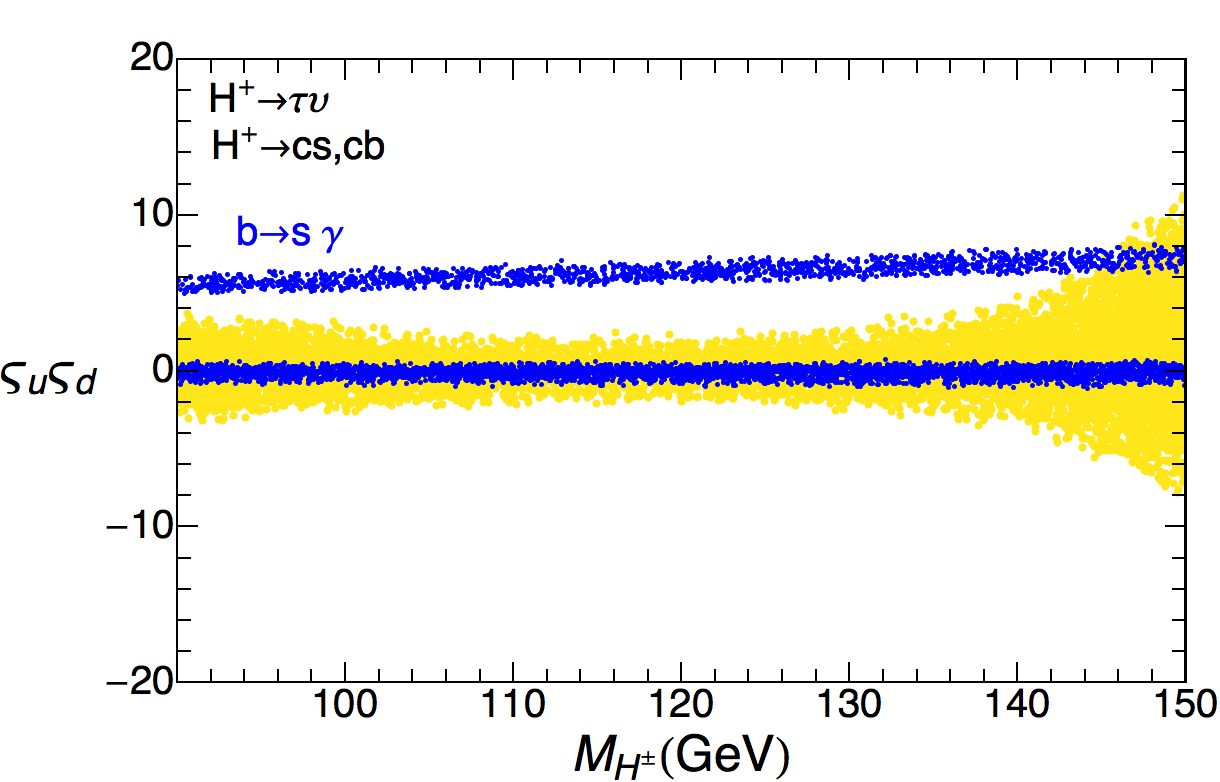}
\hskip 1.25cm
\includegraphics[width=7.3cm,height=6.3cm]{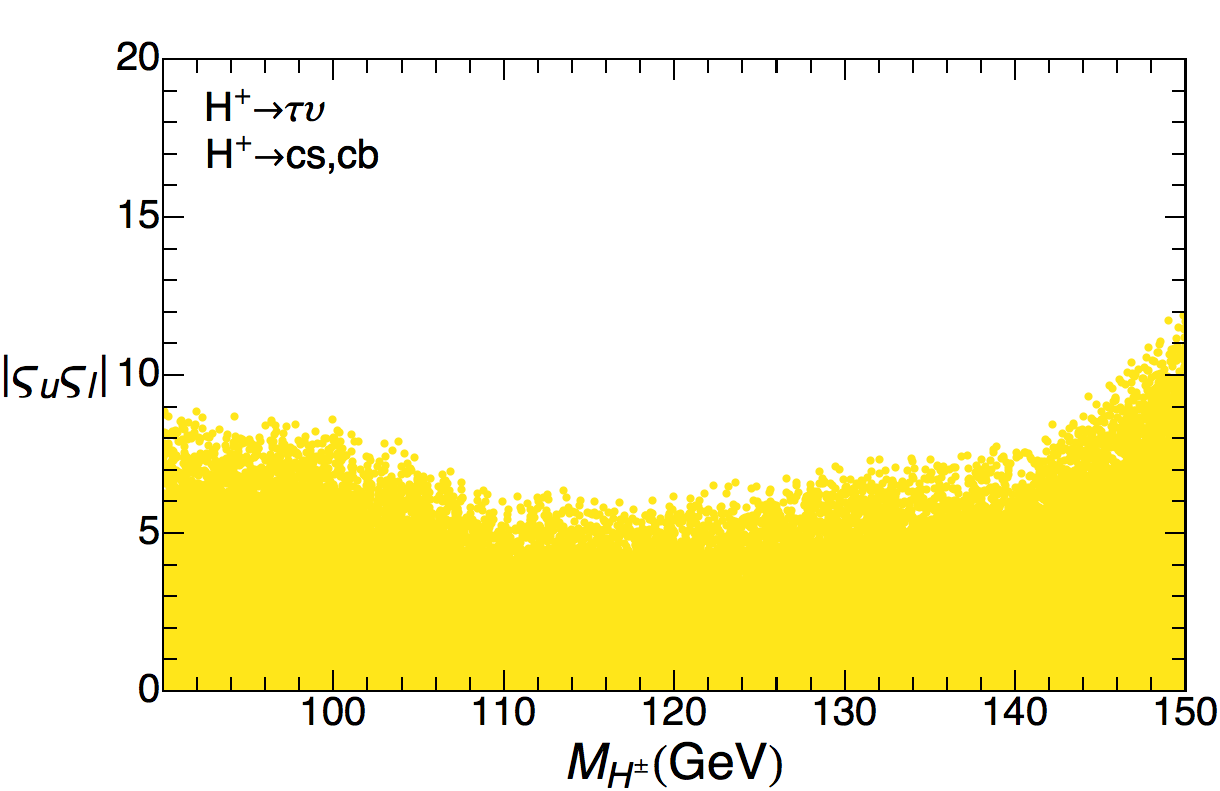}
\caption{\label{Charged_const} \it \small  Left-panel: Allowed values for $\varsigma_u \varsigma_d$ as a function of the charged Higgs mass (yellow-light) obtained from the experimental $95\%$~CL upper bounds on $\mathrm{Br}( t \rightarrow H^+ b  ) \times  \left[ \mathrm{Br}(  H^+ \rightarrow c \bar s  )  + \mathrm{Br}(  H^+ \rightarrow c \bar b  )\right]$ and $\mathrm{Br}( t \rightarrow H^+ b  ) \times  \mathrm{Br}(  H^+ \rightarrow \tau^+ \nu  )$.   Allowed values for $\varsigma_u \varsigma_d$ from $\mathrm{Br}(\bar B \rightarrow X_s \gamma)$
are shown in blue-dark.   Right-panel: Similar constraints on the combination $|\varsigma_u \varsigma_l|$ from direct charged Higgs searches.   The alignment parameters have been varied in the range $|\varsigma_u| \leq1$ and $|\varsigma_{d,l}| \leq 50$.
   }
\end{figure}

In Figure~\ref{Charged_const} we show the bounds on the A2HDM parameter space from direct searches of a light charged Higgs at the LHC.  Note that the present upper bounds on $\mathrm{Br}( t \rightarrow H^+ b  ) \times  \left[ \mathrm{Br}(  H^+ \rightarrow c \bar s  )  + \mathrm{Br}(  H^+ \rightarrow c \bar b  )\right]$ and $\mathrm{Br}( t \rightarrow H^+ b  ) \times  \mathrm{Br}(  H^+ \rightarrow \tau^+ \nu  )$ set an upper limit on $|\varsigma_u \varsigma_{l}|/M_{H^{\pm}}^2$ of $\mathcal{O}(\lesssim 10^{-3})~\text{GeV}^{-2}$.  Leptonic $B$, $D$ and $D_s$ meson decays put weaker constraints on this combination, $\varsigma_u \varsigma_l/M_{H^{\pm}}^2 \in [-0.006, 0.037] \cup [0.511, 0.535]$~GeV$^{-2}$ at $95\%$~CL~\cite{Jung:2010ik}.   Moreover an upper bound on the combination $|\varsigma_u \varsigma_d|$ is obtained from direct charged Higgs searches. Semileptonic and leptonic meson decays, on the other hand, only constrain the combinations $\varsigma_u \varsigma_l$ and $\varsigma_d \varsigma_l$~\cite{Jung:2010ik}.    For both decay rates: $\Gamma(t\to H^+ b)$ and $\Gamma(H^+\to u_i \bar{d_j},  \tau^+ \nu )$, see Eqs.~(\ref{eq::tHb}) and (\ref{eq::Huidj}), terms proportional to $ \varsigma_u \varsigma_{d}$ or $ \varsigma_u \varsigma_{l}$ are negligible.  Thus, no information on the relative sign between $\varsigma_u$ and $\varsigma_{d, l}$ is obtained.

Allowed values at $90\%$~CL from the loop-induced process $\bar{B} \rightarrow X_s \gamma$~\cite{Jung:2010ab,Jung:2012vu} on the $(M_{H^\pm}, \varsigma_u \varsigma_d )$ plane are also shown in Figure~\ref{Charged_const}.  They are given by the two narrow (blue, dark) horizontal strips. We observe that, with the exception of the small region for which $M_{H^\pm} \sim [140,150]$ GeV, the upper strip is already excluded by direct $H^\pm$ searches.  $\bar{B} \rightarrow X_s \gamma$ impose no additional constraints on the combination $(M_{H^\pm},|\varsigma_u \varsigma_l|)$. For all given points in Figure~\ref{Charged_const} we find that $|\varsigma_u|\leq 0.5$, which is fully compatible with the flavour constraints given by $R_b$ and neutral meson mixing~\cite{Jung:2010ik}.

\begin{figure}[tb]
\centering
  \includegraphics[width=7.3cm,height=6.3cm]{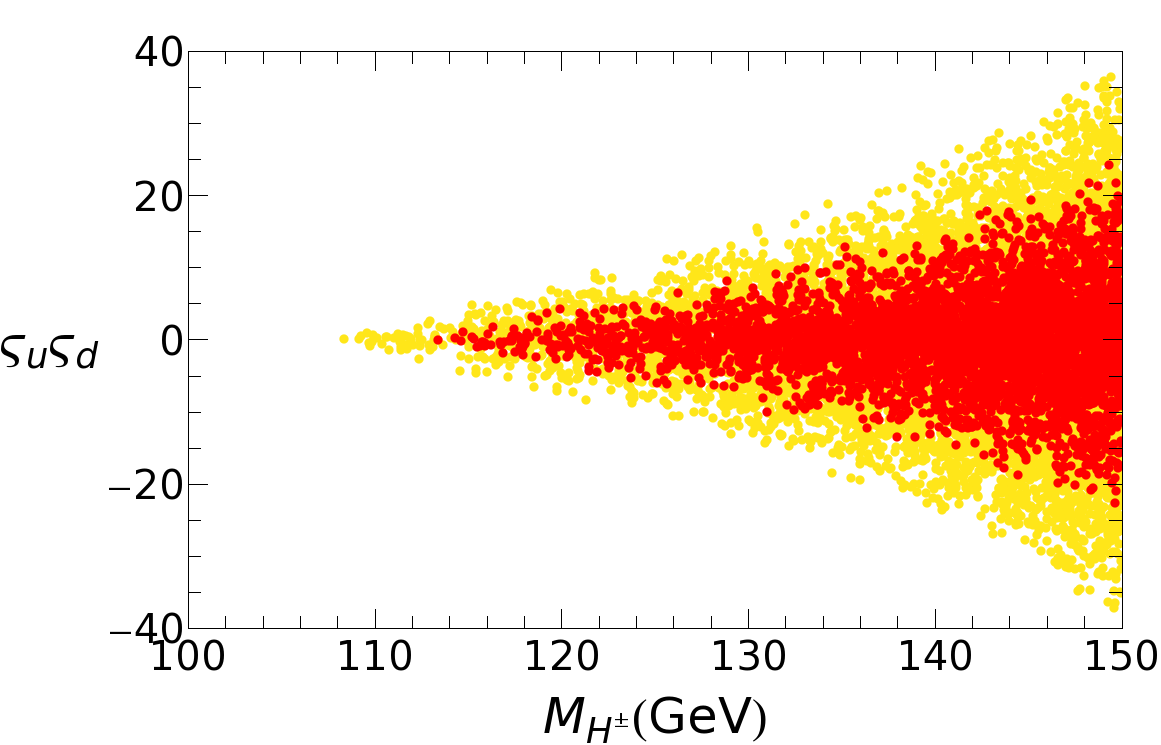}
 \!\  \!\  \!\  \!\  \!\
 \includegraphics[width=7.3cm,height=6.3cm]{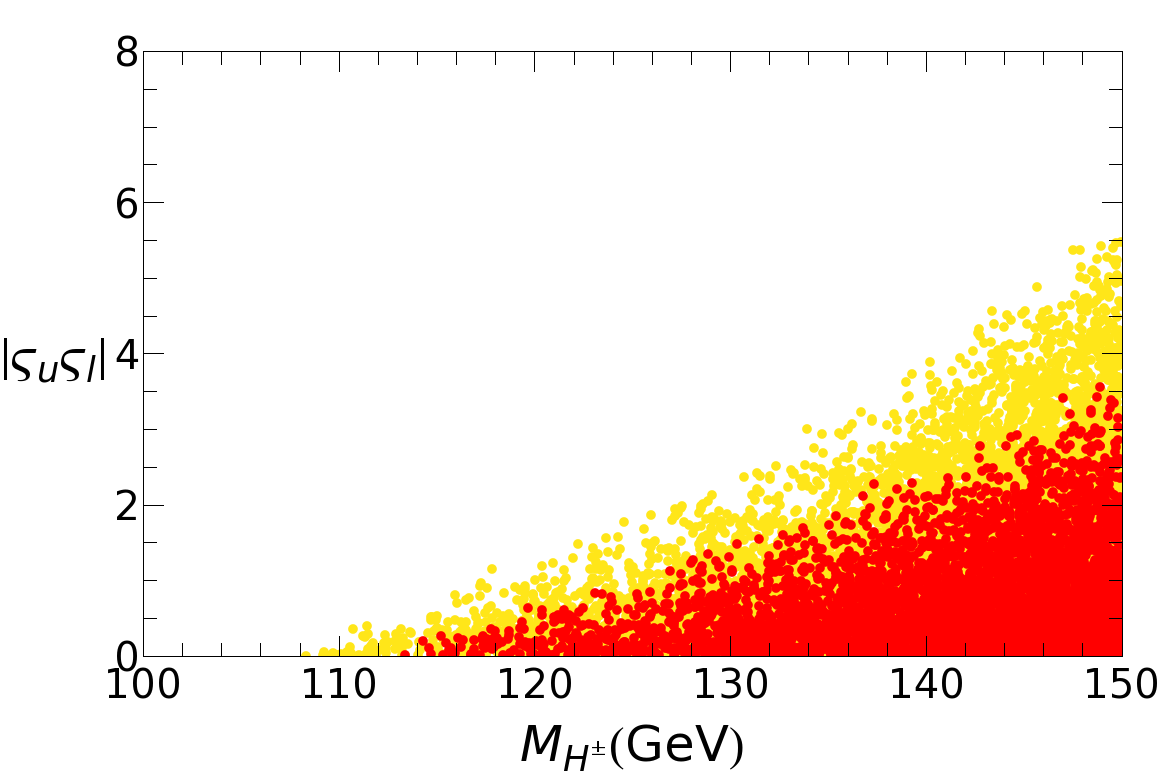}   \!\  \!\  \!\  \!\
\caption{ \it \small  Region in the $M_{H^\pm}-\varsigma_u \varsigma_d$ (left) and $M_{H^\pm}-|\varsigma_u \varsigma_l|$ (right) planes which satisfy the condition $\mathrm{Br}(H^+\to W^+b\bar{b}) > 10 \% $ (yellow, light) and $\mathrm{Br}(H^+\to W^+b\bar{b}) > 20 \% $ (red, dark).  The alignment parameters have been varied in the range $|\varsigma_u| \leq1$ and $|\varsigma_{d,l}| \leq 50$. }
\label{zuzfvitualtop}
\end{figure}

In the A2HDM, the three-body decay $H^+ \to t^*\bar{b}\to W^+b\bar{b}$ can also play an important role for a light charged Higgs when $M_{H^{\pm}} > M_W + 2 m_b $, see appendix~\ref{formulae}. This decay is normally very suppressed for a large region of the parameter space. It has been previously analyzed in Refs.~\cite{virTop::Djouadi,virTop::Wudka,virTop::Borzumati,virTop::Stirling,virTop::Bi} and it was found that it can bring a sizeable contribution to the total charged Higgs decay rate in the $\mathcal{Z}_2$ models or in the MSSM when $M_{H^\pm} > 135$--$145$ GeV, depending on the model and on the chosen value of $\tan \beta$. In the A2HDM it can bring sizeable contributions to the branching fraction, of the order of 10--20\%, already when $M_{H^\pm} \gsim 110$ GeV.  Figure~\ref{zuzfvitualtop} shows the regions satisfying the condition $\mathrm{Br}(H^+\to W^+b\bar{b}) > 10\%$ (20\%), in the planes $M_{H^\pm}-\varsigma_u \varsigma_d$ and $M_{H^\pm}-|\varsigma_u \varsigma_l|$.  There are wide regions that can bring potentially large contributions to the decay rate, and that partially overlap with the allowed regions shown in Figure~\ref{Charged_const}. If we reanalyze the previous experimental constraints from the direct charged Higgs searches by adding this channel to the total decay rate, the allowed regions stay roughly the same, however, the allowed points concentrate in the region $|\varsigma_u\varsigma_d|\lesssim 1.5$.   Thus, we conclude that experimental direct searches for a charged Higgs should be enlarged by also including this channel.

It is also worth noticing that for a fermiophobic charged Higgs, for which $\varsigma_{f=u,d,l}=0$ and hence, $H^\pm$ does not couple to fermions at tree-level, all experimental constraints are trivially satisfied.   Other production mechanisms and decay channels would have to be considered in this case to experimentally probe such scenario.

\subsection{Neutral Higgs searches} \label{nhs}
 The ATLAS and CMS collaborations have searched for additional neutral Higgs bosons up to masses of $1$~TeV in the $\varphi  \rightarrow ZZ$ and $\varphi \rightarrow WW$ channels~
\cite{ATLAS:2013nma,Chatrchyan:2013yoa}.   These searches are sensitive in principle to the heavy CP-even Higgs $H$, given that the CP-odd Higgs does not couple at tree-level with vector bosons.  Having observed no signal, they have set upper bounds on the relevant cross section $\sigma(pp \rightarrow \varphi \rightarrow VV)$, using $\sim 5$~fb$^{-1}$ and $\sim 20$~fb$^{-1}$ of collected data at $\sqrt{s}= 7$~TeV and $\sqrt{s}= 8$~TeV respectively.   Searches for neutral bosons in the leptonic final state $\tau^+ \tau^-$ with masses up to $500$~GeV have been performed by the ATLAS collaboration, using $\sim 5$~fb$^{-1}$ of collected data at $\sqrt{s}= 7$~TeV~\cite{Aad:2012cfr}.   Bounds in the $\tau^+ \tau^-$ channel have also been presented recently by the CMS collaboration, using the full $2011+2012$ dataset, for Higgs masses up to $1$~TeV~\cite{CMS:2013hja}.   These searches are sensitive to both CP-even and CP-odd Higgs bosons.  Since the CP-odd Higgs does not couple at tree-level with vector bosons, its decay branching ratios into fermions are expected to be large.    We assume in this section that the heavy scalars $H$ and $A$ cannot decay in non-SM decay channels like $H/A \rightarrow hh$; the bounds obtained here would be weaker if these decay channels were relevant.  This assumption is well justified only in certain regions of the parameter space, namely, when $M_H < 2 M_h$ or if the relevant cubic Higgs self-couplings are very small. 

\begin{figure}[tb]
\centering
\includegraphics[width=9.cm,height=6.cm]{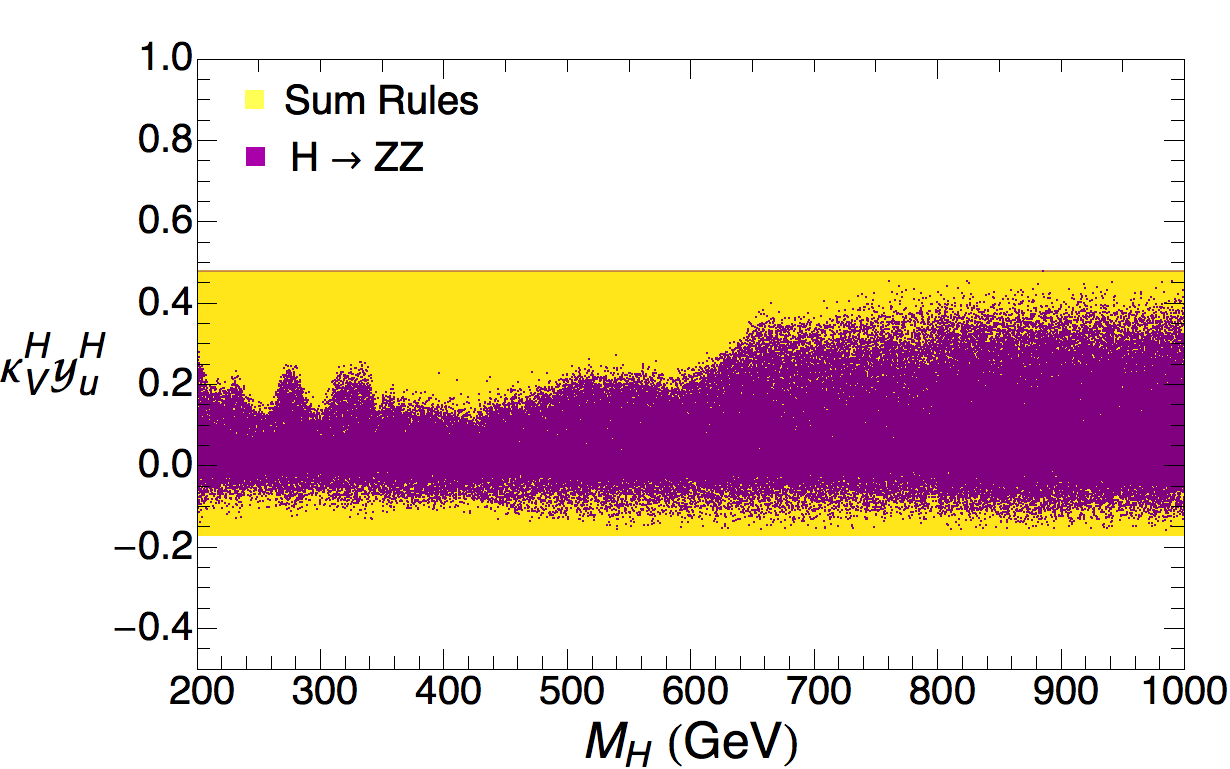}
\caption{\label{testsumrule} \it \small
Allowed values $(90\%$~CL) for the combination $\kappa_V^{H} y_{u}^{H}$ due to generic sum rules,  taking into account $h(126)$ collider data and flavour constraints (yellow-light).   Experimental limits on $\sigma(pp \rightarrow H \rightarrow ZZ)$ are also included, shrinking the allowed region to the purple-dark area.    }
\end{figure}

At present, searches for heavy scalars in the $H \rightarrow ZZ$ channel are the most sensitive, reaching $\sigma(pp \rightarrow H \rightarrow ZZ)/\sigma(pp \rightarrow H \rightarrow ZZ)_{\mathrm{SM}} \sim 10^{-1}$ for $M_{H}  \lesssim 600$~GeV.  Generic constraints on the properties of the missing 2HDM scalars can also be obtained from $h(126)$ collider data and flavour observables due to the sum rules governing the scalar couplings.     Bounds on the combination $\kappa_{V}^{H}  y_{u}^{H}$, as determined in Eq.~\eqref{Eqfs}, are shown in Figure~\ref{testsumrule} (yellow-light).     Current experimental limits on $\sigma(pp \rightarrow H \rightarrow ZZ)$ are also included in Figure~\ref{testsumrule}, reducing the allowed parameter space to the purple-dark area.   It can be observed that for heavier Higgs masses the bounds become weaker as expected.

\begin{figure}[tb]
\centering
\includegraphics[width=6.cm,height=6.cm]{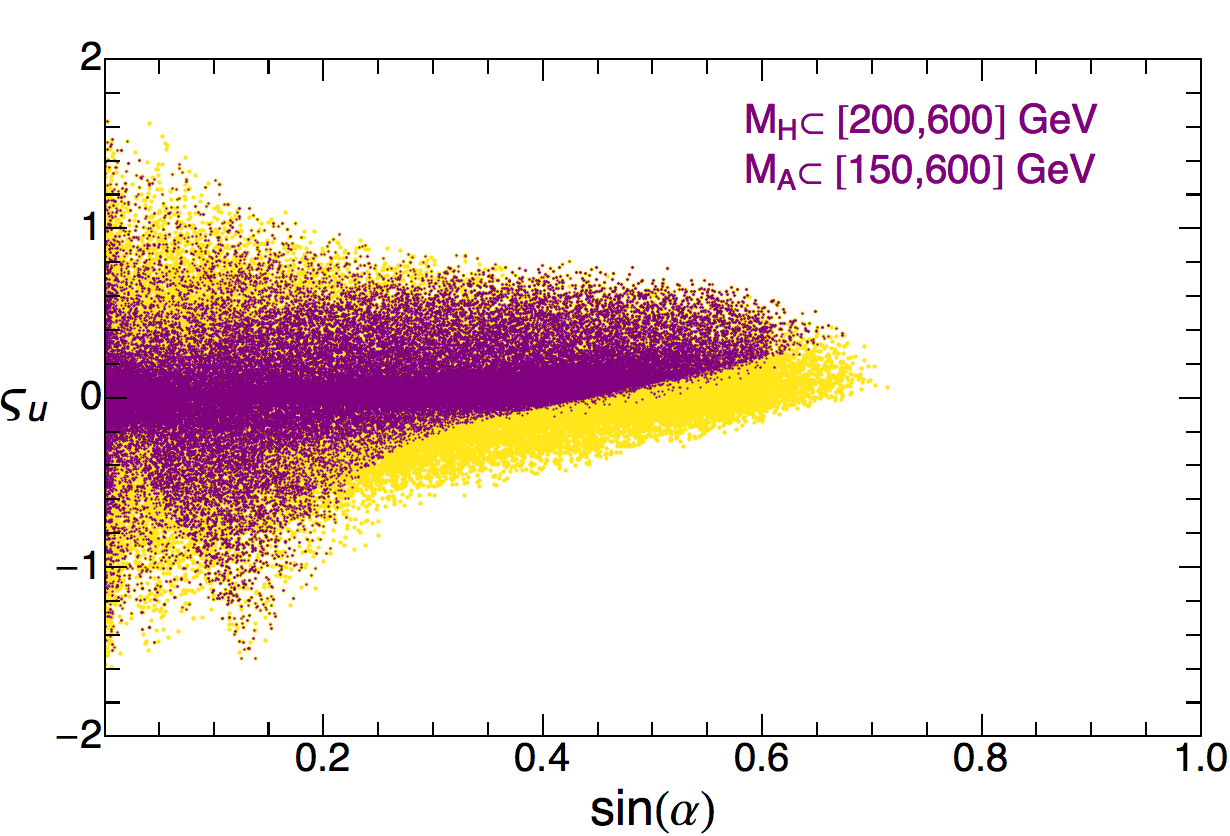}
\hskip 1.2cm
\includegraphics[width=6.cm,height=6.cm]{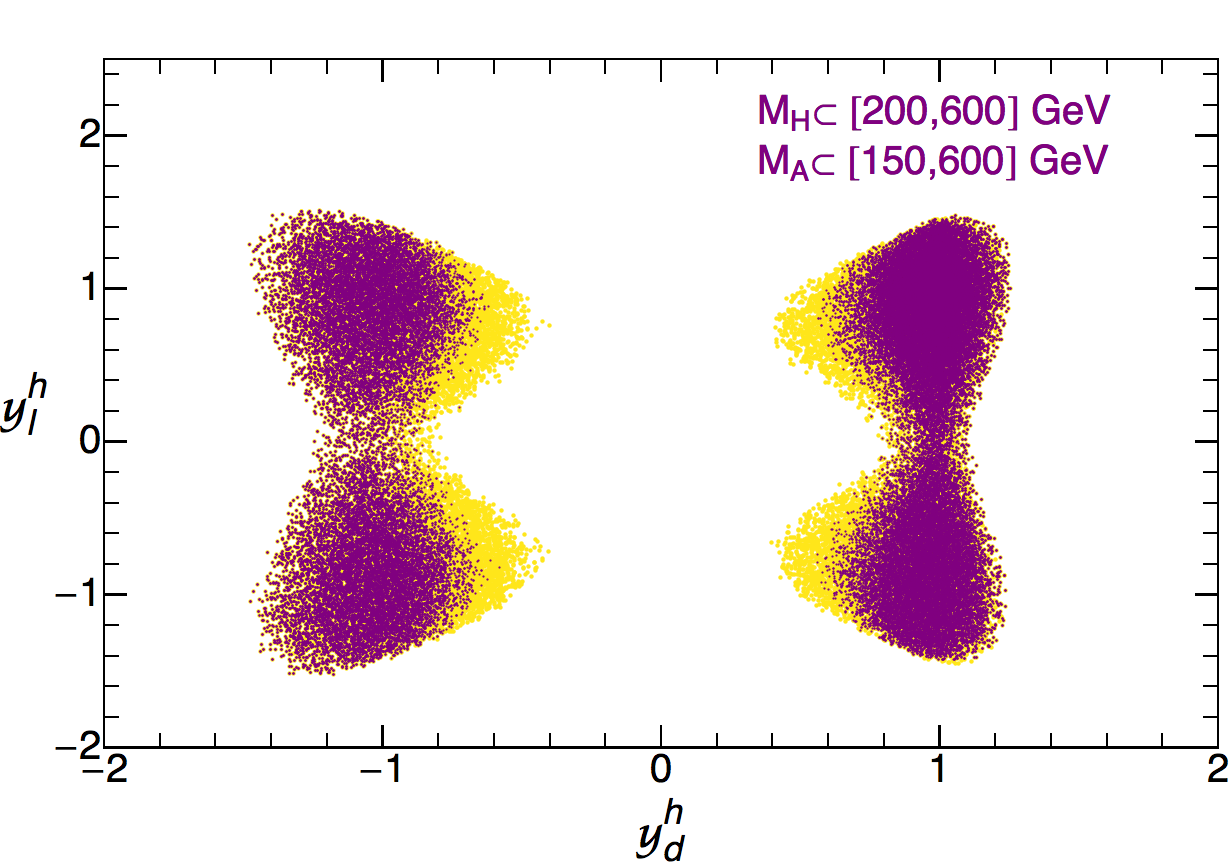}\\
\includegraphics[width=6.cm,height=6.cm]{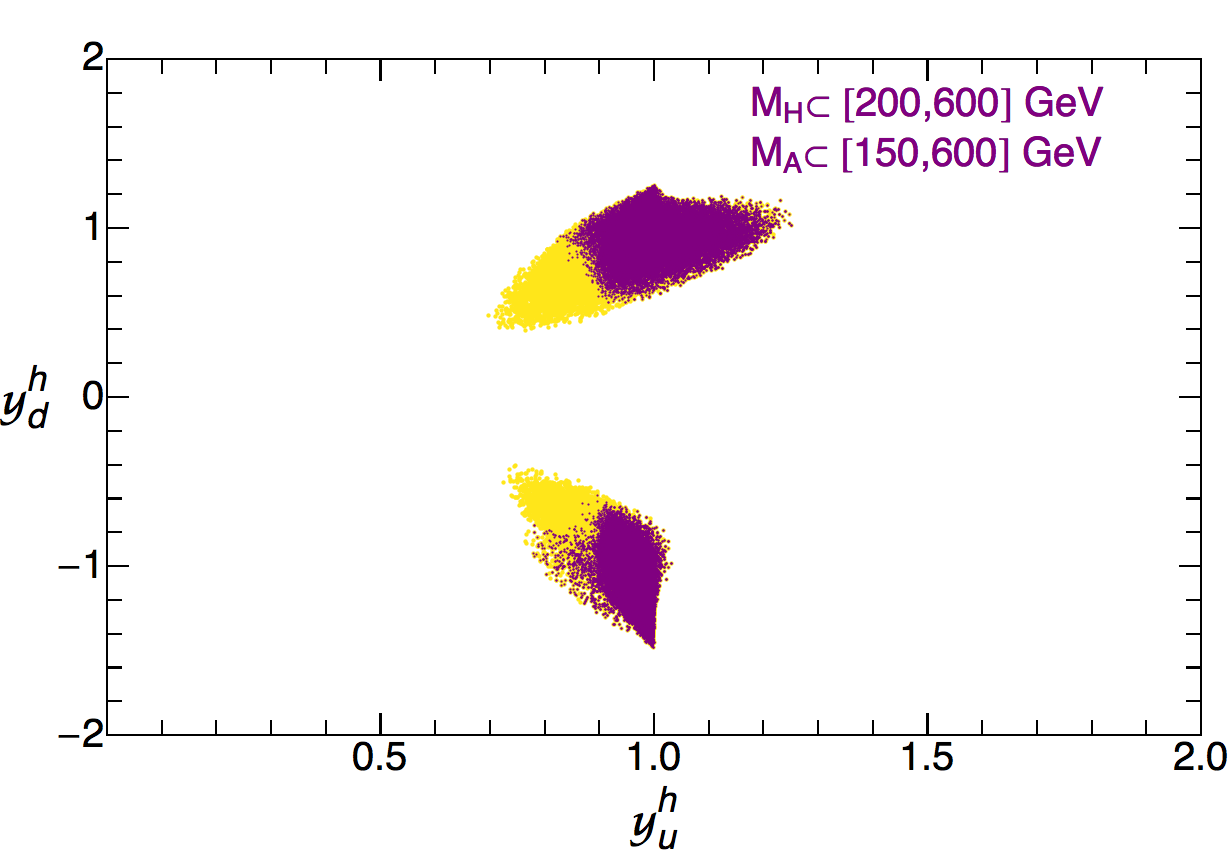}
\hskip 1.2cm
\includegraphics[width=6.cm,height=6.cm]{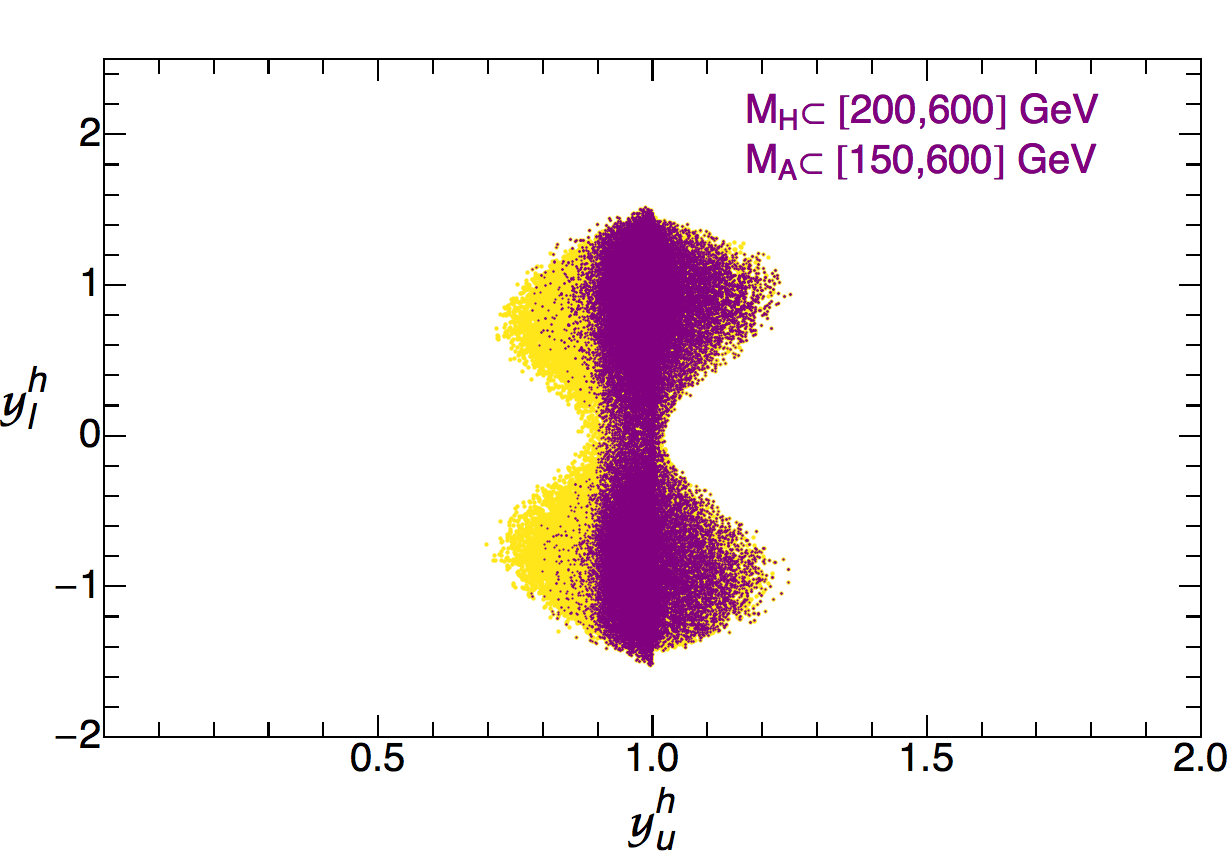}
\caption{\label{fig:RealiBII} \it \small
Allowed regions in the planes  $\sin \tilde \alpha-\varsigma_u$ (top-left), $y_d^h- y_l^h$ (top-right),  $y_u^h- y_d^h$ (bottom-left), and $y^h_u - y_l^h$ (bottom-right) at 90\%~CL, from a global fit of $h(126)$ collider data together with $R_b$ and $\mathrm{Br}(\bar B \rightarrow X_s \gamma)$, within the CP-conserving A2HDM, are shown in yellow-light.   Constraints from neutral Higgs searches at the LHC have also been included taking $M_{H} \in [200, 600]$~GeV and $M_{A} \in [150, 600]$~GeV, shrinking the allowed region to the purple-dark area, see text for details.    }
\end{figure}

To assess the impact of direct searches for additional scalars to further restrict the available parameter space of the 2HDM, we take the heavy CP-even
and CP-odd Higgses to lie in the mass ranges: $M_{H} \in [200, 600]$~GeV and $M_A \in  [150, 600]$~GeV.    Of course, a similar analysis could be performed in any other mass ranges for $H$ and $A$, or by also including constraints from collider searches of a charged Higgs.   Here, we have varied the masses of the CP-even
and CP-odd scalars
independently.   Electroweak precision data gives rise to correlations in the $M_H - M_A$ plane depending on the value of the charged Higgs mass, as shown in Figure~\ref{fig:oblique}.   At this point however, this does not have any impact on the allowed regions found in Figures~\ref{testsumrule} and \ref{fig:RealiBII}.   

In Figure~\ref{fig:RealiBII} we show the allowed regions (yellow-light) obtained in section~\ref{global}, considering the $h(126)$ collider data together with the flavour observables $R_b$ and $\mathrm{Br}(\bar B \rightarrow X_s \gamma)$.   The allowed regions get reduced when taking into account the limits from direct searches of additional scalars at the LHC (purple-dark). The most important effects are a lower bound on $y_{u}^{h}$ and a smaller allowed area in the $\varsigma_u - \sin\tilde\alpha$ plane,
which are mainly
due to the present experimental upper limits on $\sigma( pp \rightarrow H \rightarrow ZZ)$; current searches in the $\tau^+ \tau^-$ and $W^+W^-$ channels put weaker constraints.  The production cross section via gluon fusion scales as $\sigma(gg \rightarrow H)Ê\propto | y_{u}^{H} |^2 = |\sin \tilde \alpha - \varsigma_u \cos \tilde \alpha |^2$ (neglecting the contributions from other quarks which are in general subdominant). When $\sin \tilde \alpha $ is far from zero, the decay channels $H \rightarrow VV$ ($V= ZZ, W^+ W^-$) are the dominating ones, given that the fermionic couplings are not very large as the LHC and Tevatron data seem to suggest.    The production cross section $\sigma(gg \rightarrow H)$ will then grow for negative values of $\varsigma_u$, giving rise to a significant total cross section that becomes excluded by the present upper limits on $\sigma( pp \rightarrow H \rightarrow ZZ)$.

\section{The fermiophobic charged Higgs scenario \label{sec:fermiophobic}}
In the limit $\varsigma_{f=u,d,l} = 0$ the charged Higgs does not couple to fermions at tree level.  A very light fermiophobic charged Higgs, even below $80$~GeV, is perfectly allowed by data.   All bounds coming from flavour physics or direct charged Higgs searches that involve the $H^{\pm}$ couplings to fermions are naturally evaded in this case.    It is also known that when $|\kappa_{V}^{h}| = |\cos \tilde \alpha| \simeq 1$ (which is presently favoured by LHC and Tevatron data), the process $h \rightarrow 2 \gamma$ provides a unique place were non-decoupling effects can be manifest if $M_{H^{\pm}} \sim \mathcal{O}(v)$~\cite{Gunion:2002zf}.  This motivates a dedicated analysis of this scenario in light of the latest collider data.    Here we assume that the lightest CP-even state $h$ is the $126$~GeV boson and that CP is a good symmetry of the scalar sector, as in the previous section.   The scaling of the neutral Higgs couplings to vector bosons and fermions becomes equal in this limit, $y_{f}^{h} =  \kappa_{V}^{h}$, which makes this scenario very predictive in the neutral scalar sector.    The $h \rightarrow 2 \gamma$ decay width is approximately given in this case by
\be
 \frac{\Gamma( h \rightarrow \gamma \gamma)}{\Gamma( h \rightarrow \gamma \gamma)^{\mathrm{SM}}} \; \simeq \; \left(\, \kappa_{V}^{h} -  0.15 \, C_{H^{\pm}}^{h}   \right)^2   \,,
\ee
where $C_{H^{\pm}}^{h}$ encodes the charged Higgs contribution to the $h \rightarrow 2 \gamma$ decay width.  More specifically, $\mathcal{C}_{H^\pm}^{h}= v^2/(2M_{H^\pm}^2) \, \lambda_{h H^+H^-} \, \mathcal{A}(x_{H^\pm})\,$ with $x_{H^{\pm}} = 4 M_{H^{\pm}}^2/M_h^2$,  the cubic Higgs coupling is defined through $\mathcal{L}_{h H^+H^-} = - v \, \lambda_{h H^+H^-} \,h H^+H^-$ and the loop function $\mathcal{A}(x)$ is given by
\be
\mathcal{A}(x) \;=\;   - x - \frac{x^2}{4} f(x)  \,, \qquad \qquad f(x) = - 4 \,\mathrm{arcsin}^2(1/\sqrt{x})  \,.
\ee
Here we have assumed that $M_{H^{\pm}} > M_h/2 \simeq 63~$GeV so that $\mathcal{C}_{H^\pm}^{h}$ does not contain an imaginary absorptive part.  The cubic Higgs self coupling $\lambda_{h H^+H^-}$ can be expressed as a linear combination of quartic couplings of the scalar potential in the Higgs basis, see for example Ref.~\cite{Celis:2013rcs}.    To reduce the number of parameters to a minimal set, we perform a fit to $(\cos \ta, C_{H^{\pm}}^{h})$, treating $C_{H^{\pm}}^{h}$ as a free real variable.    A full scan of the scalar parameter space, taking into account electroweak precision data, vacuum stability of the potential, perturbativity and perturbative unitarity bounds, would of course give rise to non-trivial correlations between the relevant Higgs self couplings and the scalar masses. 
\begin{figure}[tb]
\centering
\includegraphics[width=7.3cm,height=7.3cm]{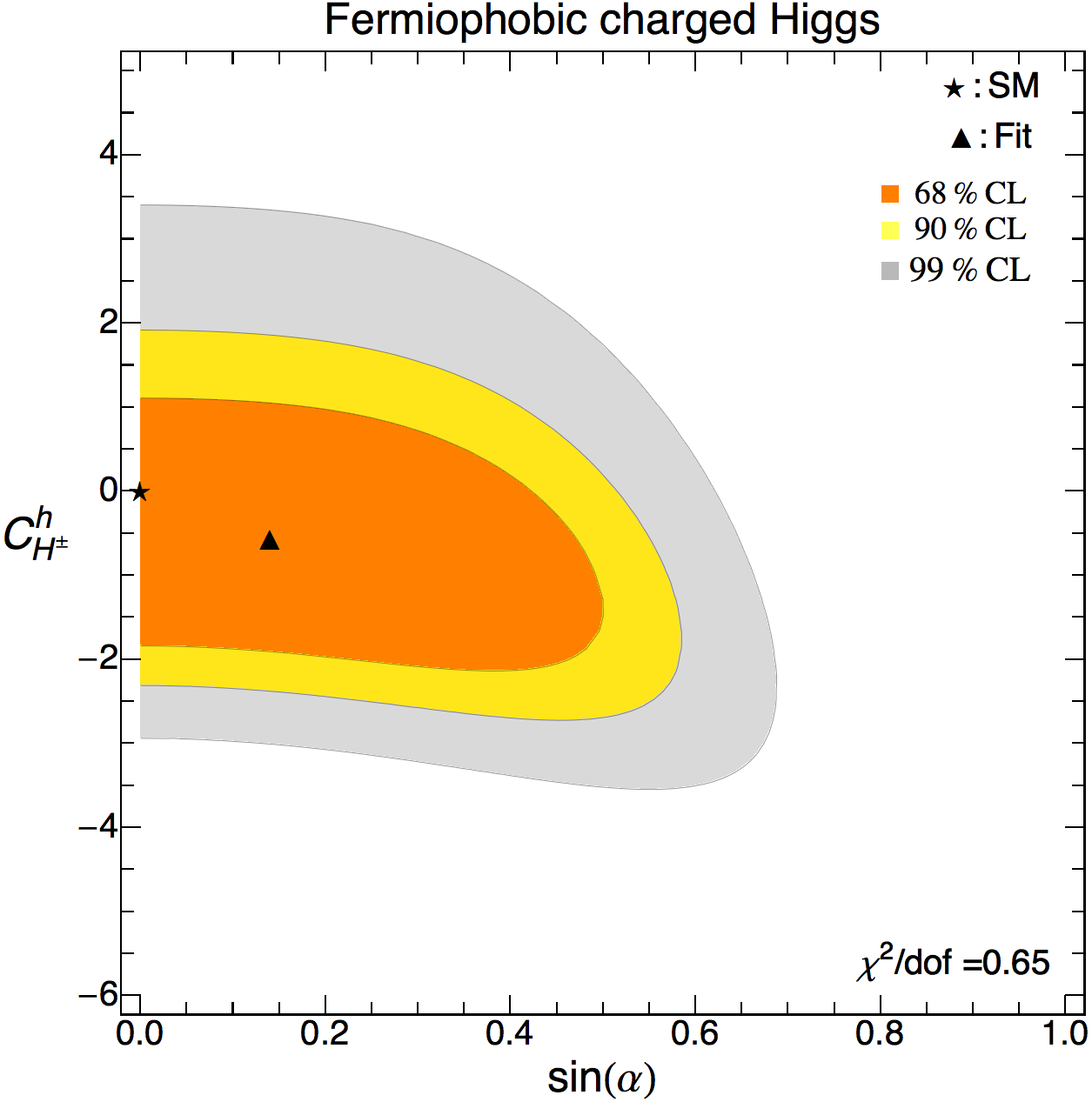}
\hskip 1.25cm
\includegraphics[width=7.3cm,height=7.3cm]{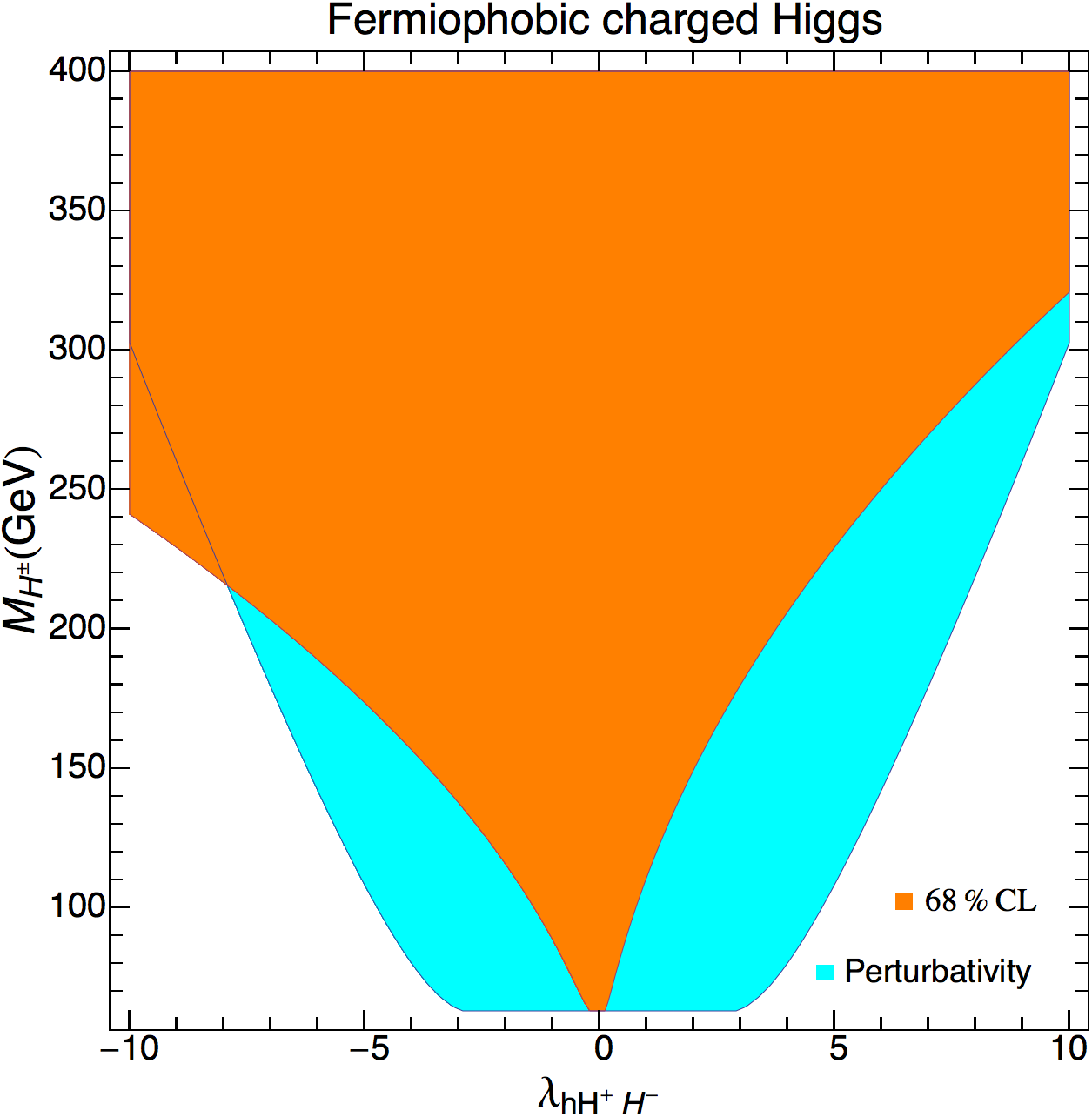}
\caption{\label{Higgs125Charged} \it \small
Allowed regions at 68\% (orange), 90\% (yellow) and 99\%~CL (grey) for a fermiophobic charged Higgs in the plane $\sin \tilde \alpha-\mathcal{C}_{H^\pm}^{h}$ (left). The right plot shows the corresponding 68\%~CL  (orange) region in the parameters $\lambda_{h H^+H^-}$ and $M_{H^\pm}$.  The region where perturbation theory remains valid is indicated in light-blue.}
\end{figure}

The best fit to the data is obtained for $(\cos \tilde \alpha, C_{H^{\pm}}^{h}) = (0.99, -0.58)$ with $\chi_{\mathrm{min}}^2/\dof  \simeq 0.65$.  In Figure~\ref{Higgs125Charged} (left) we show the allowed regions at 68\% (orange), 90\% (yellow) and 99\% (gray)~CL in the variables $( \sin \tilde \alpha, C_{H^{\pm}}^{h})$.  In the right panel of Figure~\ref{Higgs125Charged}, the resulting constraint on $C_{H^{\pm}}^{h}$ at $68\%$~CL is shown in terms of the cubic Higgs coupling $\lambda_{h H^+ H^-}$ and the charged Higgs mass $M_{H^{\pm}}$.   The perturbativity limits on the cubic Higgs coupling $h H^+ H^-$, discussed in Ref.~\cite{Celis:2013rcs}, are also indicated (light-blue).    The allowed region in the plane $(\lambda_{h H^+ H^-}, M_{H^{\pm}})$ is slightly tilted towards negative $\lambda_{h H^+ H^-}$ values, since the best fit point prefers a small negative charged Higgs contribution to the $h \rightarrow 2 \gamma$ decay amplitude.

At $90\%$~CL, we find for the Higgs signal strengths:\footnote{Higgs signal strengths refer to Higgs cross sections normalized by the SM prediction, $\mu_{X}^{\varphi} = \sigma( pp \rightarrow \varphi \rightarrow X)/\sigma( pp \rightarrow  \varphi \rightarrow X)_{\mathrm{SM}}$.} $\mu_{\bar b b}^{h} = \mu_{\bar \tau \tau}^{h} = \mu_{\bar WW, ZZ}^{h}  = \cos^2 \tilde \alpha \in [0.74,1]$ and $\mu^{h}_{\gamma \gamma}  = 1.13 \pm 0.48$.  These relations between the Higgs signal strengths hold in any of the relevant Higgs production mechanisms~\cite{Celis:2013rcs}.

Heavy Higgs boson searches in the channels $W^+W^-$ and $ZZ$ are sensitive to the gauge coupling $\kappa_V^{H}$ and to cubic scalar couplings relevant to describe possible non-SM decay channels like $H \rightarrow hh$.   In the following we assume that the later can be neglected, this implies that the analysis presented here is only valid in certain regions of the parameter space.  We find then that $\mu_{WW,ZZ}^{H} = \sin^2 \tilde \alpha \leq 0.26$ at $90\%$~CL.  Considering the current experimental limits on $\mu^H_{WW,ZZ}$~\cite{ATLAS:2013nma,Chatrchyan:2013yoa}, one can rule out a heavy CP-even Higgs in the mass range $M_{H} \in [130, 630]$~GeV when $\sin^2 \tilde \alpha = 0.26$; this bound disappears of course when $\sin \tilde \alpha \rightarrow 0$, since $H$ decouples from the vector bosons and the fermions. Associated charged Higgs production with a $W^{\pm}$ boson via neutral Higgs decays, $\varphi_j^0 \rightarrow H^{\pm} \, W^{\mp}$, with the charged Higgs decaying later to lighter neutral Higgs bosons, is a possible channel to probe the fermiophobic charged Higgs scenario.   Sum rules among the couplings $g_{\varphi_j^0 H^{\pm} W^{\mp}}$ imply that $|g_{h H^{\pm} W^{\mp}}/g_{H H^{\pm} W^{\mp}}| = |\sin \tilde \alpha/ \cos \tilde \alpha| < 0.6$ at $90\%$~CL, while $g_{A H^{\pm} W^{\mp}}$ is completely fixed by the gauge symmetry~\cite{Celis:2013rcs}. Since the charged Higgs does not decay into fermions at tree level, branching fractions for $H^{\pm} \rightarrow \varphi_j^0 \, W^{\pm}$ decays can be particularly large.

An even more restricted scenario in which the charged Higgs decouples from the fermions is given by the Inert 2HDM.  In this case a $\mathcal{Z}_2$ symmetry is imposed in the Higgs basis so that all SM fields and $\Phi_1$ are even under this symmetry while $\Phi_2  \rightarrow - \Phi_2$.  Therefore, there is no mixing between the CP-even neutral Higgs bosons $h$ and $H$. Assuming that the $h(126)$ boson corresponds to the lightest CP-even Higgs, we then have that $y_{f}^{h} = 1$ and $\cos \ta = 1$.  If furthermore one assumes that there are no open decay channels other than the SM ones,  only the diphoton channel can show a deviation from the SM due to the charged Higgs contribution.  From a global fit of this scenario to LHC and Tevatron data we obtain $C_{H^{\pm}}^{h}  \in [-1.9, 1.2]$ at $90\%$~CL ($\chi_{\mathrm{min}}^2/\mathrm{dof} \simeq 0.6$).  This can be compared with the situation before Moriond 2013 in which $C_{H^{\pm}}^{h} \in [-2.4,-0.1]$ at $90\%$~CL, driven by the excess in the diphoton signal observed at the moment~\cite{Celis:2013rcs}.   Detailed studies of the Inert 2HDM, discussing the possibility to account for the Dark Matter in the Universe, can be found in Refs.~\cite{Krawczyk:2013jta,Goudelis:2013uca,Arhrib:2013ela} and references therein.

\section{Comparison with other works}
\label{compas}
Following the discovery of the $h(126)$ boson, a large number of works have appeared, analyzing the implications of collider data within the framework of 2HDMs.  The majority of these analyses have been performed assuming NFC~\cite{Barroso:2013zxa,Grinstein:2013npa,Eberhardt:2013uba,Chen:2013rba,Craig:2013hca,Coleppa:2013dya,Shu:2013uua,Chiang:2013ixa,Krawczyk:2013jta,Goudelis:2013uca,Arhrib:2013ela,Belanger:2013xza,Enberg:2013ara,WoudaII,Chang:2013ona,Cheung:2013rva
},
thus restricting considerably the Yukawa structure of the model and the phenomenological possibilities.
The ATLAS and CMS collaborations were initially observing a significant excess in the diphoton channel.  The most natural explanation for such excess was a large charged Higgs contribution to the $h \rightarrow \gamma \gamma$ decay amplitude, other alternatives being usually in conflict with flavour constraints or perturbativity bounds, see Ref.~\cite{Celis:2013rcs} and references therein.    The situation has changed drastically after Moriond 2013, given that the CMS collaboration now reports a diphoton signal that is no longer enhanced.
The main message that can be extracted from recent analyses is that current collider data can be accommodated very well in the SM; the addition of a second Higgs doublet does not improve in a significant way the agreement with the data.  Important constraints start to appear for 2HDMs with NFC, restricting them to lie closer to the SM-limit.

Considerable work has also been done recently to analyze the future prospects at the LHC, as well as in possible future machines, to detect additional Higgs bosons within 2HDMs.   Compared with the vast literature on the subject before the $h(126)$ discovery, information about the $h(126)$ boson properties is now being included in these analyses.
Phenomenological studies within 2HDMs with NFC, relevant for the search of additional scalars, have been done in Refs.~\cite{Grinstein:2013npa,Eberhardt:2013uba,Craig:2013hca,Chen:2013rba,Chiang:2013ixa,Enberg:2013ara,WoudaII,Brownson:2013lka,Chen:2013qda,Asner:2013psa,Arhrib:2013oia}.  Promising production mechanisms and decay channels have been pointed out in these works.  In particular, if the $h(126)$ couplings are found to be very close to those of the SM,  searches for heavy neutral Higgs bosons in the channels $\gamma \gamma$ or $\tau^+ \tau^-$ become particularly relevant~\cite{Craig:2013hca}.  It could also be possible that heavy Higgs bosons decay mostly into the lightest state $h$, assumed to be the $h(126)$ boson.  In this case, $h$ production via heavy Higgs decays could be the way to detect these heavy states~\cite{Arhrib:2013oia}.   Some possibilities for this scenario are $H \rightarrow hh$, $A \rightarrow Z h$, and $H^{\pm} \rightarrow W^{\pm} h$.  In any case,  the non-observation of additional Higgs bosons will provide complementary information, together with direct measurements of the $h(126)$ boson properties, to restrict the parameter space of 2HDMs.

The experimental collaborations have also shown interest to search for signatures of extended Higgs sectors at the LHC, beyond the usually tested minimal supersymmetric scenarios.   The ATLAS collaboration, for example, has released a search for a heavy CP-even Higgs boson in the $H \rightarrow WW\rightarrow e \nu\mu\nu$ channel within the types I and II 2HDMs, in the mass range $[135, 300]$~GeV, using $13$~fb$^{-1}$ of data at $\sqrt{s} =  8$~TeV center of mass energy~\cite{ATLAS:2013zla}.    The CMS collaboration, on the other hand, has analyzed the future prospects in the search for heavy neutral Higgs bosons at the LHC.    The analysis was performed in the channels $H \rightarrow ZZ \rightarrow 4 \ell ~(\ell = e, \mu)$ and $ A \rightarrow Z h \rightarrow \ell \ell bb$, assuming an integrated luminosity of 3000~fb$^{-1}$ at $\sqrt{s} = 14$~TeV center of mass energy~\cite{CMS:2013dga}.     On the experimental side, the main challenge seems to account for the large number of free parameters present in the 2HDM, even in the more restricted versions with NFC.  On the theoretical side there is still a lot of work to be done to be able to start a precision study of these more general
extended Higgs sectors.   Theoretical predictions for cross-sections and branching ratios, taking into account relevant electroweak and QCD corrections, as well as its implementation in standard tools will be of utmost importance as experimental data becomes more precise, see for example Refs.~\cite{Eriksson:2009ws,Harlander:2012pb,Harlander:2013mla,Englert:2013vua} for some relevant works in this direction.

In this work, we have focused on the possibility of performing a more general analysis of collider data within the framework of 2HDMs, without resorting to any symmetry in the Yukawa sector as is done in the different scenarios with NFC.  The A2HDM provides a rich Yukawa structure that includes all the different 2HDMs with a $\mathcal{Z}_2$ symmetry as particular limits while, at the same time, suppresses flavour changing transitions in low-energy systems to acceptable levels~\cite{Pich:2009sp,Jung:2010ik,Jung:2010ab,Jung:2012vu}.     First studies of the $h(126)$ boson data within the A2HDM, in the CP-conserving limit, were performed in Refs.~\cite{Cervero:2012cx,Altmannshofer:2012ar,Bai:2012ex,Celis:2013rcs} and more recently in Refs.~\cite{Barger:2013ofa,Lopez-Val:2013yba}.   The role of new sources of CP-violation beyond the CKM-phase present in the A2HDM were also discussed in Ref.~\cite{Celis:2013rcs}; we will consider this possibility in more detail in a future work.       The main problem one has to face in this approach is
the larger number of free parameters,
compared with the NFC models.    On the other hand, one is able to perform in this way non-biased analyses of the scalar sector of the 2HDM, without imposing symmetries which at first hand might seem ad-hoc.  We have shown for example how generic sum-rules governing the scalar couplings provide a direct connection between the $h(126)$ properties and those of the missing scalars, see Eq.~\eqref{eq:SumRulesP}.

A comprehensive analysis of current $h(126)$ data within extended Higgs sectors
has been recently performed in Ref.~\cite{Lopez-Val:2013yba}, including
comparisons between the A2HDM and different $\mathcal{Z}_2$ 2HDMs.
Also of relevance in this work, is a discussion of the effect of quantum corrections in relation to high-precision studies of the Higgs sector.  In Ref.~\cite{Barger:2013ofa}, emphasis was given on an estimation of the future sensitivity that can be achieved at a high-luminosity LHC, a linear electron-positron collider and a muon collider, making the relevant comparisons between the A2HDM and the different NFC scenarios.
A discussion of possible phenomenological strategies to test the 2HDM has been done recently in the Higgs basis~\cite{Asner:2013psa}, following the basis independent methods developed in Ref.~\cite{Davidson:2005cw}.

Information about the $h(126)$ boson properties is crucial for making simplifying assumptions and reducing the number of relevant variables, in order to perform a viable scan of the 2HDM parameter space at the LHC or at future colliders.
In this work, we have analyzed the current data,
keeping only a minimal set of parameters that are of relevance while capturing the rich phenomenology provided by the Yukawa structure of the A2HDM.

\section{Summary}
\label{sec:summary}
We have studied the implications of LHC and Tevatron data, after the first LHC shutdown, for CP-conserving 2HDMs, assuming that the $h(126)$ boson corresponds to the lightest CP-even state of the scalar spectrum.  The phenomenological analysis has been done within the general framework of the A2HDM, which contains as particular limits all different 2HDMs based on $\mathcal{Z}_2$ symmetries.
Interesting bounds on the properties of the additional Higgs bosons of the model can be extracted, due to the existence of sum rules relating the different scalar couplings.

The $h(126)$ coupling to vector bosons is found to be very close to the SM limit, implying an upper bound on the heavy CP-even Higgs coupling to vector bosons: $|\kappa_{V}^{H}|  < 0.6$ at $90\%$~CL.   Other bounds on the couplings of the missing neutral scalars have been summarized in Eq.~(\ref{Eqfs}).  The flipped-sign solution for the top-quark Yukawa coupling, which was preferred by the fit before Moriond 2013 in order to explain the excess in the $2 \gamma$ channel~\cite{Celis:2013rcs}, is now found to be excluded at $90\%$~CL.    A sign degeneracy in the determination of the bottom and tau Yukawa couplings however remains.

We have discussed the role of flavour physics constraints, electroweak precision observables and LHC searches for additional scalars to further restrict the parameter space.   Some results of our analysis can be pointed out.  Loop-induced processes $(Z \rightarrow \bar b b$ and $\bar B \rightarrow X_s \gamma)$ set important constraints on the quark Yukawa couplings, $y_{u}^{h}$ and $y_{d}^{h}$, for charged Higgs masses below $500$~GeV.   Also, heavy Higgs searches in the $ZZ$ channel put significant limits on the up-type quark Yukawa coupling $y_{u}^{h}$.  Regarding direct charged Higgs searches at colliders, decays of the charged Higgs into a $c \bar b$ pair and three-body decays $H^+ \rightarrow t^* \bar b \rightarrow W^+ b \bar b$, can have sizable decay rates in some regions of the allowed parameter space.   Future searches for a light charged Higgs at the LHC in hadronic final states should take
these possibilities into account, perhaps through the implementation of b-tagging techniques as suggested in Ref.~\cite{Akeroyd:2012yg}.

The fermiophobic charged-Higgs scenario has been discussed in light of current experimental data.  Though this is a particular limit of the A2HDM, it deserved a separate analysis for different reasons.  A very light fermiophobic charged Higgs boson can give unusually large contributions to the $h \rightarrow  \gamma \gamma$ amplitude.    Another reason is that in this case many simple relations arise between the properties of the neutral Higgs bosons, making this scenario particularly predictive when analyzing the searches for additional Higgs bosons at the LHC.    We find that current data still allow for very light charged scalars and sizable contributions from a charged Higgs to the $h \rightarrow 2 \gamma$ amplitude.

\begin{appendix}

\section{Useful formulae for a light charged Higgs}
\label{formulae}
A light charged Higgs with $M_{H^{\pm}} < m_t  + m_b$ can be produced at the LHC via top-quark decays.  The relevant partial decay widths are given by
\begin{align}
\Gamma(t \rightarrow W^+ b) \; &= \;  \frac{g^2  \!\ |V_{tb}|^2}{64  \!\   \pi  \!\ m_t^3}  \;\, \lambda^{1/2}(m_t^2,m_b^2,M_W^2)\; \left( m_t^2 + m_b^2 + \frac{(m_t^2-m_b^2)^2}{M_W^2} - 2 M_W^2\right) \,, \\[2ex]
\Gamma(t \rightarrow H^+ b) \; &= \;   \frac{|V_{tb}|^2}{16 \pi m_t^3 v^2}\;\, \lambda^{1/2}(m_t^2, m_b^2, M_{H^{\pm}}^2 )   \;  \Bigl[ (  m_t^2 + m_b^2 - M_{H^{\pm}}^2  )(     m_b^2 |\varsigma_d|^2 + m_t^2 |\varsigma_u|^2  )\quad  \nonumber \\
&\hskip 6cm\mbox{}
- 4 m_b^2 m_t^2 \,\mathrm{Re}(  \varsigma_d \varsigma_u^* ) \Bigr]\,,\label{eq::tHb}
\end{align}
with $\lambda(x,y,z) = x^2 + y^2 + z^2 - 2(x y + x z + y z )$ and $g= 2 M_W/v$.    QCD vertex corrections to $t \rightarrow H^{\pm} b$ and $t \rightarrow W^{\pm} b$ cancel to a large extent in $\mathrm{Br}(t \rightarrow H^{\pm} b)$~\cite{Li:1990cp}.   The charged Higgs decays into quarks and leptons are described in the A2HDM by the following expressions:
\begin{align} \label{eq::Huidj}
\Gamma(H^+ \rightarrow l^+ \nu_{l}) \; &=\; \frac{m_{l}^2}{8 \pi v^2 } \, \left(  1- \frac{m_{l}^2}{M^2_{H^{\pm}}} \right)^2   \, M_{H^{\pm}}\,|\varsigma_l|^2   \,,   \nonumber   \\[0.3cm]
\Gamma(H^+ \rightarrow  u_i \bar{d_j} ) \; &=\;
\frac{N_C  \, |V_{ij}|^2}{8 \pi v^2 M_{H^{\pm}}^3 }\;
\lambda^{1/2}( M_{H^{\pm}}^2, m_{u_i}^2, m_{d_j}^2) \;
\left(1 + \frac{17}{3}\, \frac{\alpha_s(M_{H^{\pm}}) }{\pi} \right)
\nonumber \\ &\mbox{}\; \times\;\,
\Bigl[ ( M_{H^{\pm}}^2  - m_{u_i}^2  - m_{d_j}^2 )     (|\varsigma_d|^2 m_{d_j}^2 + | \varsigma_u |^2 m_{u_i}^2   )
+ 4 m_{u_i}^2 \, m_{d_j}^2 \, \mathrm{Re}(\varsigma_d \,\varsigma_u^*)     \Bigr]   \,,
\end{align}
where $N_C$ is the number of colours.  Running $\overline{ \mathrm{MS}}$ quark masses entering in these expressions are evaluated at the scale $M_{H^{\pm}}$, and the leading QCD vertex correction to $H^{+} \rightarrow u \bar d$ has been taken into account~\cite{Braaten:1980yq}.

 \begin{figure}[th]
\centering
\includegraphics[scale=0.5]{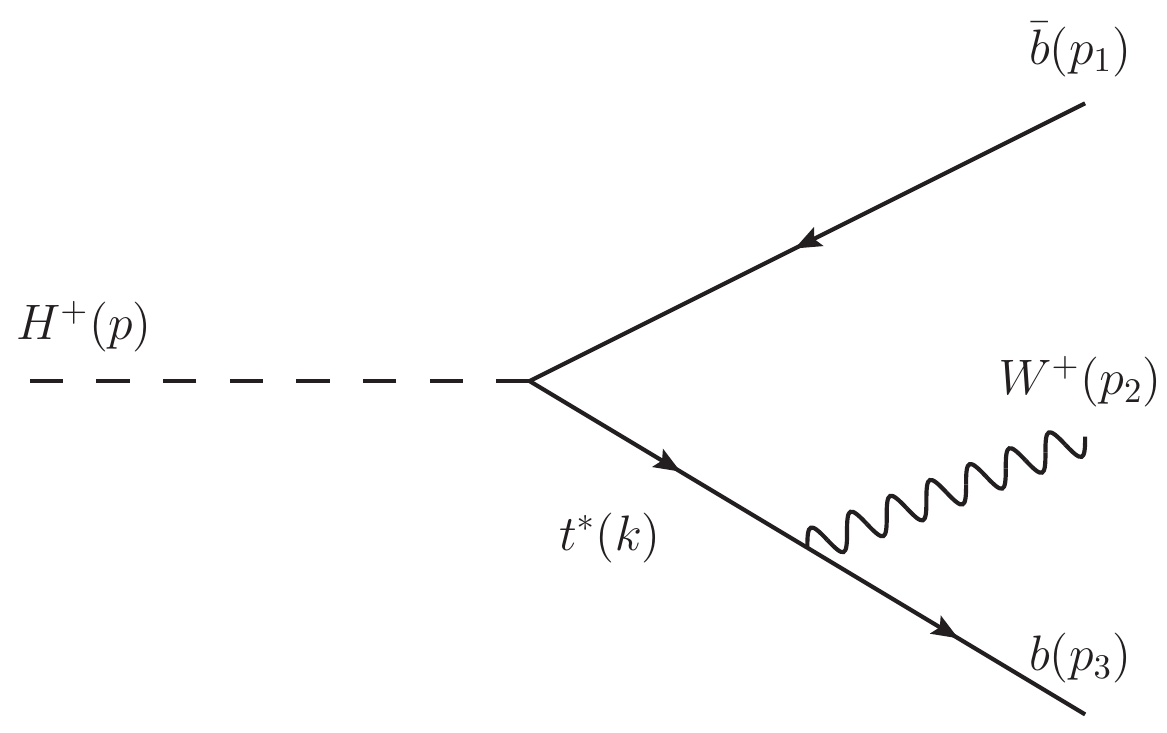}
\caption{  \it \small  Feynman diagram for the three-body charged Higgs decay $H^+ \to t^*\bar{b}\to W^+b\bar{b}$.}
\label{virtualTop}
\end{figure}

When the charged Higgs mass satisfies $M_{H^{\pm}} > M_W + 2 m_b $, three-body decays of the charged Higgs mediated by a virtual top quark can be relevant, see Figure~\ref{virtualTop}.   The decay width for $H^+ \to t^*\bar{b}\to W^+b\bar{b}$ is given in the A2HDM by
\be
\Gamma(H^\pm \to t^*\bar{b}\to W^+b\bar{b})\; =\; \frac{N_C \ g^2 |V_{tb}|^4 }{128\pi^3 M_{H^\pm}^3 M_W^2v^2}\; \int ds_{23}\int ds_{13}\;\, \frac{G(s_{23},s_{13})}{[s_{23}-m_t^2]^2} \,,
\ee
where
\begin{align}
G(s_{23},s_{13})\; &=\; \left[ M_W^2 (p_1 p_3) + 2\, (p_2 p_3)(p_1 p_2)\right]\,\left[|\varsigma_u|^2 \,  m_t^4 - |\varsigma_d|^2  \, m_b^2  \!\  k^2\right]  \notag \\ &+\; \left[M_W^2  m_b^2  \,  (p_3 k) + 2m_b^2 \, (p_2 p_3)(p_2 k)\right]\,\left[ 2\, |\varsigma_d|^2 \, (p_1 k) + 2m_t^2 \, \mathrm{Re}(\varsigma_u\varsigma_d^*)\right] \,,
\end{align}
\noindent with:
\begin{align}
k&=p_2+p_3 \, ,  \qquad\qquad  k^2=s_{23}  \, , \qquad\qquad
(p_1p_3)=\frac{1}{2}(s_{13}-2m_b^2) \,,\notag \\
(p_2p_3)&=\frac{1}{2}(s_{23}-M_W^2 -m_b^2)  \, , \qquad\qquad
(p_1p_2)=\frac{1}{2}(M_{H^\pm}^2+m_b^2-s_{23}-s_{13})\,.
\end{align}
\noindent The integration limits are:
\begin{align}
s_{23}^{\text{min}}\; =\; \frac{1}{4s_{13}}\; \left\{ (M_{H^\pm}^2 - M_W^2)^2 - \left[\lambda^{1/2}(M_{H^\pm}^2,s_{13},M_W^2) + \lambda^{1/2}(s_{13},m_b^2,m_b^2)\right]^2  \right\}  \,,\notag \\
s_{23}^{\text{max}}\; = \;\frac{1}{4s_{13}}\; \left\{ (M_{H^\pm}^2 - M_W^2)^2 - \left[\lambda^{1/2}(M_{H^\pm}^2,s_{13},M_W^2) - \lambda^{1/2}(s_{13},m_b^2,m_b^2)\right]^2  \right\} \,,
\end{align}
with
\begin{align}
4m_b^2 \leqslant s_{13} \leqslant (M_{H^\pm} -M_W)^2 \,.
\end{align}

\section{Statistical treatment and experimental data}
\label{statistical}
The experimental $h(126)$ data used in the fit can be found in Tables~\ref{tab:dataATLAS} and \ref{tab:dataTevatron}; experimental uncertainties are assumed to be Gaussian.   To obtain the preferred values for the parameters of the A2HDM we build a global $\chi^2$ function.    For some channels the correlation coefficient $\rho$ between different production modes can be estimated from the $68\%$~CL contours provided by the experimental collaborations,  assuming that the $\Delta \chi^2 = \chi^2 - \chi^2_{\text{min}}$ is well described by a bivariate normal distribution.   This information is taken into account in the fit.  The $68\%$ and $90\%$ one-dimensional confidence level (CL) intervals are given by $\Delta\chi^2 = 1$ and $2.71$, respectively.   Two-dimensional $68\%$ and $90\%$ CL intervals are given by $\Delta\chi^2 = 2.30$ and $4.31$, respectively.

\begin{table}[t]\begin{center}
\caption{\it \small Experimental data from the ATLAS and CMS collaborations at $\sqrt{s} = 7+8$~TeV.  }
\vspace{0.2cm}
\begin{tabular}{|c||c|c||c|c|c|}
\hline
Channel & $\hat \mu$ (ATLAS) &     Comment   & $\hat \mu$   (CMS) &      Comment  \\
\hline \hline
$bb(\text{VH})$ &  $ 0.25 \pm 0.65 $ & Ref.~\cite{Aad:2013wqa} & $1.0 \pm 0.5$   &  Ref.~\cite{Chatrchyan:2013lba} \\ \hline
$\tau \tau(\text{ggF})$ &  $2.19 \pm 2.2 $  &   $\rho= -0.50$  &  $0.68 \pm 1.05 $   &  $\rho = -0.5$   \\
$\tau \tau(\text{VBF + VH})$ &  $-0.31 \pm 1.25 $  &  Ref.~\cite{Aad:2013wqa}    &  $ 1.57  \pm 1.13  $     & Ref.~\cite{Chatrchyan:2013lba}   \\ \hline
$WW(\text{ggF})$ &  $0.79 \pm 0.52 $  &    $\rho=-0.2$   &  $0.76 \pm 0.35 $   &  $\rho = -0.3$  \\
$WW(\text{VBF+VH})$ &  $ 1.6 \pm 1.25 $  &  Ref.~\cite{Aad:2013wqa}  &  $0.24 \pm 1.14 $   &  Ref.~\cite{Chatrchyan:2013lba} \\  \hline
$ZZ(\text{incl.})$ &  $ 1.5 \pm 0.4 $  &  Ref.~\cite{Aad:2013wqa} &  $0.92 \pm 0.28 $    &    Ref.\cite{Chatrchyan:2013lba}   \\ \hline
$\gamma \gamma(\text{ggF})$ &  $1.6 \pm 0.6 $  & $\rho=-0.3$  &  $ 0.47 \pm 0.49 $  & $\rho = -0.6$ \\
$\gamma \gamma(\text{VBF+VH})$ &  $1.76 \pm 1.28 $ & Ref.~\cite{Aad:2013wqa}  &  $ 1.6 \pm 1.14 $    & Ref.~\cite{Chatrchyan:2013lba} \\
\hline
\end{tabular}
\label{tab:dataATLAS}
\end{center}\end{table}

\begin{table}[t]\begin{center}
\caption{\it \small Experimental data from CDF and D\O~ at $\sqrt{s} = 1.96$~TeV.  }
\vspace{0.2cm}
\begin{tabular}{|c||c|c| }
\hline
Channel & $\hat \mu$ &    Comment  \\
\hline \hline
$bb(\text{VH})$ &  $1.59 \pm 0.71  $  & Ref.~\cite{Aaltonen:2012qt} \\ \hline
$\tau \tau(\text{incl.})$ &  $1.7 \pm 2.0  $  &  Ref.~\cite{Aaltonen:2012qt} \\ \hline
$WW(\text{incl.})$ &  $0.94 \pm 0.84$  &  Ref.~\cite{Aaltonen:2012qt} \\  \hline
$\gamma \gamma(\text{incl.})$ &  $5.97 \pm 3.25 $  & Ref.~\cite{Aaltonen:2012qt} \\
\hline
\end{tabular}
\label{tab:dataTevatron}
\end{center}\end{table}

Regarding the flavour observables considered in this work, we use the latest $\bar B \rightarrow X_s \gamma$ experimental measurement, $\mathrm{Br}( \bar B \rightarrow X_s \gamma)_{E_0 > 1.6~\text{GeV}} =  (3.41 \pm 0.22) \times 10^{-4}  $~\cite{Lees:2012ym}.  The theoretical prediction of this quantity is obtained following Ref.~\cite{Misiak:2006ab}.  The calculation of $R_b$ within 2HDMs was detailed in Ref.~\cite{Degrassi:2010ne}; the experimental value is $R_b = \Gamma(Z \rightarrow \bar b b)/\Gamma(Z \rightarrow \text{hadrons}) = 0.21629 \pm 0.00066$~\cite{Alcaraz:2009jr}.

\end{appendix}

\section*{Acknowledgements}
We thank Xin-Qiang Li and Martin Jung for fruitful collaborations related to the flavour constraints on the A2HDM.  We also acknowledge useful discussions with Luca Fiorini and Emilie Passemar regarding the experimental data. This work has been supported in part by the Spanish
Government and ERDF funds from the EU Commission
[Grants FPA2011-23778 and CSD2007-00042
(Consolider Project CPAN)] and by Generalitat
Valenciana under Grant No. PROMETEOII/2013/007. The work of A.C. is supported by the Spanish Ministry MECD through the FPU grant AP2010-0308.  The work of V.I. is supported by the Spanish Ministry MEC through the FPI grant BES-2012-054676.


\begin{thebibliography}{100}



\bibitem{Aad:2012tfa} ATLAS Collaboration,
  Phys.\ Lett.\ B {\bf 716} (2012) 1
  [arXiv:1207.7214 [hep-ex]].

\bibitem{Aad:2013wqa} ATLAS Collaboration,
  Phys.\ Lett.\ B {\bf 726} (2013) 88
  [arXiv:1307.1427 [hep-ex]];
%
  ATLAS-CONF-2013-079 (July 19, 2013);
%
  ATLAS-CONF-2013-034 (March 13, 2013);
  David L\'opez Mateos talk at EPS 2013 for the ATLAS collaboration.

\bibitem{Chatrchyan:2012ufa} CMS Collaboration,
  Phys.\ Lett.\ B {\bf 716} (2012) 30
  [arXiv:1207.7235 [hep-ex]].

\bibitem{Chatrchyan:2013lba} CMS Collaboration,
  JHEP {\bf 06} (2013) 081
  [arXiv:1303.4571 [hep-ex]];
%
  CMS-PAS-HIG-13-005 (April 17, 2013).

\bibitem{Aaltonen:2012qt} CDF and D0 Collaborations,
  Phys.\ Rev.\ Lett.\  {\bf 109} (2012) 071804
  [arXiv:1207.6436 [hep-ex]];
%
  Phys.\ Rev.\ D {\bf 88} (2013) 052014
  [arXiv:1303.6346 [hep-ex]].

\bibitem{Aad:2013xqa} ATLAS Collaboration,
  Phys.\ Lett.\ B {\bf 726} (2013) 120
  [arXiv:1307.1432 [hep-ex]].


\bibitem{Chatrchyan:2012jja} CMS Collaboration,
  Phys.\ Rev.\ Lett.\  {\bf 110} (2013) 081803
  [arXiv:1212.6639 [hep-ex]].

\bibitem{D0-6387} D0 Collaboration, D0 Note 6387-CONF (July 22, 2013).





\bibitem{Cheung:2013kla}
  K.~Cheung, J.~S.~Lee and P.~-Y.~Tseng,
  JHEP {\bf 1305} (2013) 134
  [arXiv:1302.3794 [hep-ph]];
  J.~Ellis and T.~You,
  JHEP {\bf 1306} (2013) 103
  [arXiv:1303.3879 [hep-ph]];
  A.~Falkowski, F.~Riva and A.~Urbano,
  JHEP {\bf 1311} (2013) 111
  [arXiv:1303.1812 [hep-ph]];
  P.~P.~Giardino, K.~Kannike, I.~Masina, M.~Raidal and A.~Strumia,
  arXiv:1303.3570 [hep-ph].

\bibitem{LHCP2013} A. Pich,
   arXiv:1307.7700.

\bibitem{Gunion:1989we}
  J.~F.~Gunion, H.~E.~Haber, G.~L.~Kane and S.~Dawson,
  Front.\ Phys.\  {\bf 80} (2000) 1;
  G.~C.~Branco, P.~M.~Ferreira, L.~Lavoura, M.~N.~Rebelo, M.~Sher and J.~P.~Silva,
  Phys.\ Rept.\  {\bf 516} (2012) 1
  [arXiv:1106.0034 [hep-ph]].






\bibitem{Barroso:2013zxa}
A.~Barroso, P.~M.~Ferreira, R.~Santos, M.~Sher and J.~P.~Silva,
  arXiv:1304.5225 [hep-ph].


\bibitem{Grinstein:2013npa}
  B.~Grinstein and P.~Uttayarat,
  JHEP {\bf 1306} (2013) 094
  [arXiv:1304.0028 [hep-ph]].

\bibitem{Eberhardt:2013uba}
  O.~Eberhardt, U.~Nierste and M.~Wiebusch,
  JHEP {\bf{1307}} (2013) 118
  [arXiv:1305.1649 [hep-ph]].



\bibitem{Chen:2013rba}
  C.~-Y.~Chen, S.~Dawson and M.~Sher,
  Phys.\ Rev.\ D {\bf 88} (2013) 015018
  [arXiv:1305.1624 [hep-ph]].


\bibitem{Craig:2013hca}
  N.~Craig, J.~Galloway and S.~Thomas,
  arXiv:1305.2424 [hep-ph].



\bibitem{Coleppa:2013dya}
  B.~Coleppa, F.~Kling and S.~Su,
  arXiv:1305.0002 [hep-ph].



\bibitem{Shu:2013uua}
  J.~Shu and Y.~Zhang,
  Phys.\ Rev.\ Lett.\  {\bf 111} (2013) 091801
  [arXiv:1304.0773 [hep-ph]].




\bibitem{Chiang:2013ixa}
  C.~-W.~Chiang and K.~Yagyu,
  JHEP {\bf 1307} (2013) 160
  [arXiv:1303.0168 [hep-ph]].



\bibitem{Krawczyk:2013jta}
  M.~Krawczyk, D.~Sokolowska, P.~Swaczyna and B.~Swiezewska,
  JHEP {\bf 1309} (2013) 055
  [arXiv:1305.6266 [hep-ph]].



\bibitem{Goudelis:2013uca}
  A.~Goudelis, B.~Herrmann and O.~St\aa l,
  JHEP {\bf 1309} (2013) 106
  [arXiv:1303.3010 [hep-ph]].

\bibitem{Arhrib:2013ela}
  A.~Arhrib, Y.~-L.~S.~Tsai, Q.~Yuan and T.~-C.~Yuan,
  arXiv:1310.0358 [hep-ph].





\bibitem{Belanger:2013xza}
  G.~Belanger, B.~Dumont, U.~Ellwanger, J.~F.~Gunion and S.~Kraml,
  Phys.\ Rev.\ D {\bf 88} (2013) 075008
  [arXiv:1306.2941 [hep-ph]].


\bibitem{Enberg:2013ara}
  R.~Enberg, J.~Rathsman and G.~Wouda,
  JHEP {\bf 1308} (2013) 079
  [arXiv:1304.1714 [hep-ph]].

\bibitem{WoudaII}
  R.~Enberg, J.~Rathsman and G.~Wouda,
  arXiv:1311.4367 [hep-ph].


\bibitem{Chang:2013ona}
  S.~Chang, S.~K.~Kang, J.~-P.~Lee, K.~Y.~Lee, S.~C.~Park and J.~Song,
  arXiv:1310.3374 [hep-ph].



\bibitem{Cheung:2013rva}
  K.~Cheung, J.~S.~Lee and P.~-Y.~Tseng,
  arXiv:1310.3937 [hep-ph].



\bibitem{Pich:2009sp}
  A.~Pich and P.~Tuz\'on,
  Phys.\ Rev.\ D {\bf 80} (2009) 091702
  [arXiv:0908.1554 [hep-ph]].


\bibitem{Cervero:2012cx}
  E.~Cervero and J.~-M.~Gerard,
  Phys.\ Lett.\ B {\bf 712} (2012) 255
  [arXiv:1202.1973 [hep-ph]].

\bibitem{Altmannshofer:2012ar}
  W.~Altmannshofer, S.~Gori and G.~D.~Kribs,
  Phys.\ Rev.\ D {\bf 86} (2012) 115009
  [arXiv:1210.2465 [hep-ph]].

\bibitem{Bai:2012ex}
 Y.~Bai, V.~Barger, L.~L.~Everett and G.~Shaughnessy,
   Phys.\ Rev.\ D {\bf 87} (2013) 115013
  [arXiv:1210.4922 [hep-ph]].

\bibitem{Celis:2013rcs}
  A.~Celis, V.~Ilisie and A.~Pich,
   JHEP {\bf 1307} (2013) 053
  [arXiv:1302.4022 [hep-ph]].


\bibitem{Barger:2013ofa}
  V.~Barger, L.~L.~Everett, H.~E.~Logan and G.~Shaughnessy,
  arXiv:1308.0052 [hep-ph].


\bibitem{Lopez-Val:2013yba}
  D.~Lopez-Val, T.~Plehn and M.~Rauch,
  JHEP {\bf 1310} (2013) 134
  [arXiv:1308.1979 [hep-ph]].




\bibitem{Ilisie:2013cxa}
  V.~Ilisie,
  arXiv:1310.0931 [hep-ph].


\bibitem{Gunion:1990kf}
  J.~F.~Gunion, H.~E.~Haber and J.~Wudka,
  Phys.\ Rev.\ D {\bf 43} (1991) 904.


\bibitem{Grzadkowski:1999wj}
  B.~Grzadkowski, J.~F.~Gunion and J.~Kalinowski,
  Phys.\ Lett.\ B {\bf 480} (2000) 287
  [hep-ph/0001093].




\bibitem{Ginzburg:2004vp}
  I.~F.~Ginzburg and M.~Krawczyk,
  Phys.\ Rev.\ D {\bf 72} (2005) 115013
  [hep-ph/0408011].




\bibitem{Davidson:2005cw}
  S.~Davidson and H.~E.~Haber,
  Phys.\ Rev.\ D {\bf 72}, 035004 (2005)
  [Erratum-ibid.\ D {\bf 72}, 099902 (2005)]
  [hep-ph/0504050];
  H.~E.~Haber and D.~O'Neil,
  Phys.\ Rev.\ D {\bf 74} (2006) 015018
  [hep-ph/0602242].




\bibitem{Gunion:2002zf}
  J.~F.~Gunion and H.~E.~Haber,
  Phys.\ Rev.\ D {\bf 67} (2003) 075019
  [hep-ph/0207010].


\bibitem{Asner:2013psa}
  D.~M.~Asner, T.~Barklow, C.~Calancha, K.~Fujii, N.~Graf, H.~E.~Haber, A.~Ishikawa and S.~Kanemura {\it et al.},
  arXiv:1310.0763 [hep-ph].



\bibitem{Carena:2013ooa}
  M.~Carena, I.~Low, N.~R.~Shah and C.~E.~M.~Wagner,
  arXiv:1310.2248 [hep-ph].




\bibitem{He:2001tp}
  H.~-J.~He, N.~Polonsky and S.~-f.~Su,
  Phys.\ Rev.\ D {\bf 64} (2001) 053004
  [hep-ph/0102144];
  W.~Grimus, L.~Lavoura, O.~M.~Ogreid and P.~Osland,
  Nucl.\ Phys.\ B {\bf 801} (2008) 81
  [arXiv:0802.4353 [hep-ph]];
  H.~E.~Haber and D.~O'Neil,
  Phys.\ Rev.\ D {\bf 83} (2011) 055017
  [arXiv:1011.6188 [hep-ph]].





\bibitem{Maalampi:1991fb}
  J.~Maalampi, J.~Sirkka and I.~Vilja,
  Phys.\ Lett.\ B {\bf 265} (1991) 371;
  S.~Kanemura, T.~Kubota and E.~Takasugi,
  Phys.\ Lett.\ B {\bf 313} (1993) 155
  [hep-ph/9303263];
  A.~G.~Akeroyd, A.~Arhrib and E.~-M.~Naimi,
  Phys.\ Lett.\ B {\bf 490} (2000) 119
  [hep-ph/0006035];
  I.~F.~Ginzburg and I.~P.~Ivanov,
  Phys.\ Rev.\ D {\bf 72} (2005) 115010
  [hep-ph/0508020];
  P.~Osland, P.~N.~Pandita and L.~Selbuz,
  Phys.\ Rev.\ D {\bf 78} (2008) 015003
  [arXiv:0802.0060 [hep-ph]].






\bibitem{Jung:2010ik}
  M.~Jung, A.~Pich and P.~Tuz\'on,
  JHEP {\bf 1011} (2010) 003
  [arXiv:1006.0470 [hep-ph]].

\bibitem{Jung:2010ab}
  M.~Jung, A.~Pich and P.~Tuz\'on,
  Phys.\ Rev.\ D {\bf 83} (2011) 074011
  [arXiv:1011.5154 [hep-ph]].


\bibitem{Jung:2012vu}
  M.~Jung, X.~-Q.~Li and A.~Pich,
  JHEP {\bf 1210} (2012) 063
  [arXiv:1208.1251 [hep-ph]].




\bibitem{Hermann:2012fc}
  T.~Hermann, M.~Misiak and M.~Steinhauser,
  JHEP {\bf 1211} (2012) 036
  [arXiv:1208.2788 [hep-ph]].





\bibitem{Celis:2012dk}
  A.~Celis, M.~Jung, X.~-Q.~Li and A.~Pich,
  JHEP {\bf 1301} (2013) 054
  [arXiv:1210.8443 [hep-ph]].


\bibitem{Lees:2012xj}
  J.~P.~Lees {\it et al.}  [BaBar Collaboration],
  Phys.\ Rev.\ Lett.\  {\bf 109} (2012) 101802
  [arXiv:1205.5442 [hep-ex]].





\bibitem{Farina:2012xp}
  M.~Farina, C.~Grojean, F.~Maltoni, E.~Salvioni and A.~Thamm,
  JHEP {\bf 1305} (2013) 022
  [arXiv:1211.3736 [hep-ph]].

\bibitem{Cheung:2008zh}
  K.~Cheung, C.~-W.~Chiang and T.~-C.~Yuan,
  Phys.\ Rev.\ D {\bf 78} (2008) 051701
  [arXiv:0803.2661 [hep-ph]].


\bibitem{Lee:1977eg}
  B.~W.~Lee, C.~Quigg and H.~B.~Thacker,
  Phys.\ Rev.\ D {\bf 16} (1977) 1519.




\bibitem{Abbiendi:2013hk} ALEPH, DELPHI, L3 and OPAL Collaborations,
  Eur.\ Phys.\ J.\ C {\bf 73} (2013) 2463
  [arXiv:1301.6065 [hep-ex]].



\bibitem{Aad:2012tj} ATLAS Collaboration,
  JHEP {\bf 1206} (2012) 039
  [arXiv:1204.2760 [hep-ex]];
   ATLAS-CONF-2013-090 (August 25, 2013).

\bibitem{Chatrchyan:2012vca} CMS Collaboration,
  JHEP {\bf 1207} (2012) 143
  [arXiv:1205.5736 [hep-ex]].




\bibitem{Aad:2013hla} ATLAS Collaboration,
  Eur.\ Phys.\ J.\ C {\bf 73} (2013) 2465
  [arXiv:1302.3694 [hep-ex]].

\bibitem{Akeroyd:2012yg}
  A.~G.~Akeroyd, S.~Moretti and J.~Hern\'andez-S\'anchez,
  Phys.\ Rev.\ D {\bf 85} (2012) 115002
  [arXiv:1203.5769 [hep-ph]].





  \bibitem{virTop::Djouadi}
  A.~Djouadi, J.~Kalinowski and P.~M.~Zerwas,
  Z.\ Phys.\ C {\bf 70} (1996) 435
  [hep-ph/9511342].



  \bibitem{virTop::Wudka}
  E.~Ma, D. P.~Roy and J.~Wudka,
  Phys.\ Rev.\ Lett.\  {\bf 80} (1998) 1162
  [hep-ph/9710447].


\bibitem{virTop::Borzumati}
  F.~Borzumati and A.~Djouadi,
  Phys.\ Lett.\ B {\bf 549} (2002) 170
  [hep-ph/9806301].


\bibitem{virTop::Stirling}
  S.~Moretti and W.~J.~Stirling,
  Phys.\ Lett.\ B {\bf 347} (1995) 291
   [Erratum-ibid.\ B {\bf 366} (1996) 451]
  [hep-ph/9412209, hep-ph/9511351].


\bibitem{virTop::Bi}
  X.~-J.~Bi, Y.~-B.~Dai and X.~-Y.~Qi,
  Phys.\ Rev.\ D {\bf 61} (2000) 015002
  [hep-ph/9907326].






\bibitem{ATLAS:2013nma} ATLAS Collaboration,
  ATLAS-CONF-2013-013;
ATLAS-CONF-2013-067.



\bibitem{Chatrchyan:2013yoa} CMS Collaboration,
  Eur.\ Phys.\ J.\ C {\bf 73} (2013) 2469
  [arXiv:1304.0213 [hep-ex]];
 CMS-HIG-12-024 (July 24, 2013).


\bibitem{Aad:2012cfr} ATLAS Collaboration,
  JHEP {\bf 1302} (2013) 095
  [arXiv:1211.6956 [hep-ex]].


\bibitem{CMS:2013hja}
  CMS Collaboration,
  CMS-PAS-HIG-13-021 (November 1, 2013).




\bibitem{Chen:2013qda}
  C.~-Y.~Chen,
  arXiv:1308.3487 [hep-ph].


\bibitem{Brownson:2013lka}
  E.~Brownson, N.~Craig, U.~Heintz, G.~Kukartsev, M.~Narain, N.~Parashar and J.~Stupak,
  arXiv:1308.6334 [hep-ex].

\bibitem{Arhrib:2013oia}
  A.~Arhrib, P.~M.~Ferreira and R.~Santos,
  arXiv:1311.1520 [hep-ph].



\bibitem{ATLAS:2013zla}
  ATLAS Collaboration,
  ATLAS-CONF-2013-027 (March 10, 2013).

\bibitem{CMS:2013dga}
  CMS Collaboration,
  CMS-PAS-FTR-13-024 (October 9, 2013).


\bibitem{Eriksson:2009ws}
  D.~Eriksson, J.~Rathsman and O.~St\aa l,
  Comput.\ Phys.\ Commun.\  {\bf 181} (2010) 189
  [arXiv:0902.0851 [hep-ph]].

\bibitem{Harlander:2012pb}
  R.~V.~Harlander, S.~Liebler and H.~Mantler,
  Computer Physics Communications {\bf 184} (2013) 1605
  [arXiv:1212.3249 [hep-ph]].

\bibitem{Harlander:2013mla}
  R.~V.~Harlander, S.~Liebler and T.~Zirke,
  arXiv:1307.8122 [hep-ph].

\bibitem{Englert:2013vua}
  C.~Englert, M.~McCullough and M.~Spannowsky,
  arXiv:1310.4828 [hep-ph].




\bibitem{Li:1990cp}
  C.~S.~Li and T.~C.~Yuan,
  Phys.\ Rev.\ D {\bf 42} (1990) 3088
   [Erratum-ibid.\ D {\bf 47} (1993) 2156];
  A.~Czarnecki and S.~Davidson,
  Phys.\ Rev.\ D {\bf 48} (1993) 4183
  [hep-ph/9301237].




\bibitem{Braaten:1980yq}
  E.~Braaten and J.~P.~Leveille,
  Phys.\ Rev.\ D {\bf 22} (1980) 715;
  M.~Drees and K.~-i.~Hikasa,
  Phys.\ Lett.\ B {\bf 240} (1990) 455
   [Erratum-ibid.\ B {\bf 262} (1991) 497].




\bibitem{Lees:2012ym} BaBar Collaboration,
  Phys.\ Rev.\ Lett.\  {\bf 109} (2012) 191801
  [arXiv:1207.2690 [hep-ex]];
  Phys.\ Rev.\ D {\bf 86} (2012) 112008
  [arXiv:1207.5772 [hep-ex]].




\bibitem{Misiak:2006ab}
  M.~Misiak and M.~Steinhauser,
  Nucl.\ Phys.\ B {\bf 764} (2007) 62
  [hep-ph/0609241].






\bibitem{Degrassi:2010ne}
  G.~Degrassi and P.~Slavich,
  Phys.\ Rev.\ D {\bf 81} (2010) 075001
  [arXiv:1002.1071 [hep-ph]].



\bibitem{Alcaraz:2009jr} ALEPH, CDF, D0, DELPHI, L3, OPAL and SLD Collaborations,
  arXiv:0911.2604 [hep-ex].




\end{thebibliography}
\end{document}